\documentclass[showpacs,amsmath,amssymb,floatfix]{revtex4}

\usepackage{dcolumn}
\usepackage{amsmath,amssymb}
\usepackage{bm}
\usepackage{chngpage}
\usepackage{epsfig}
\usepackage{slashed}
\usepackage{dsfont}

\def\br{{\bf r}}

\newcommand{\nn}{\nonumber}

\begin{document}
\title{Linear response peculiarity of  a two--dimensional Dirac electron gas at weak scattering}
\author{Andreas Sinner and Klaus Ziegler}
\affiliation{Institut f\"ur Physik, Universit\"at Augsburg, D-86135 Augsburg, Germany}
\date{\today}

\begin{abstract}

The conductivity of an electron gas can be alternatively calculated either from the current--current 
or from the density--density correlation function. Here, we compare these two 
frequently used formulations of the Kubo formula for the two--dimensional Dirac 
electron gas by direct evaluations for several special cases. Assuming the presence of 
weak disorder we investigate perturbatively both formulas at and away from the Dirac point. 
While to zeroth order in the disorder amplitude both formulations give identical results, 
with some very strong assumptions though, they show significant discrepancies already 
in first order. At half filling we evaluate all second order diagrams. Virtually none 
of the topologically identical diagrams yield the same corrections for both formulations. 
We conclude that a direct comparison of conductivities of disordered system calculated 
in both formulas is not possible. The density--density correlation function is preferable since it can be linked to diffusion
via the Einstein relation.

\end{abstract}
\maketitle

\section{Introduction}

In this paper we address the observation that different versions of the 
Kubo conductivity formula do not always lead to identical results. Without claiming the
generality of this statement, we concentrate on the special case of a weakly disordered two--dimensional 
Dirac electron gas. 
On the one hand, there is a formulation in terms of the current--current
correlation function, which represents the core object in the theory of weak--localization in 
disordered metals in Refs.~[\onlinecite{Gorkov1979,Altshuler1980,Hikami1980,Altshuler1981,Ando2002,McCann2006,Suzuura2002,Khveshchenko2006}]. 
On the other hand, there is a formulation in terms of the density--density 
correlation function, which is preferably used in the theory of Anderson localization, 
Refs.~[\onlinecite{Wegner1979,Wegner1980,McKane1981,Hikami1981,Efetov1983,Fradkin1986,Wegner1989}]. Although both 
formulations are obtained starting from the same initial point, the assumptions that need to be made in between seems rather different. However, it is not the purpose of this 
paper to present a detailed discussion of these assumptions. This problem may go beyond the 
well--known fact that different limiting processes do not commute for the Kubo formula 
(cf. Ref.~[\onlinecite{ziegler07}]). Rather, we want to point out the 
discrepancies at the level of practical calculations.

The conductivity at the Fermi energy $\mu$, derived from linear response within the Kubo formalism, can be expressed 
either by the (properly normalized) current-current correlation function as
\begin{equation}
\label{eq:Kubo1}
\bar\sigma^{}_{} = 2{\rm Tr}\sum^{}_r~j^{}_n\left[\delta^{}_\eta(H-\mu)\right]^{}_{0r}
j^{}_n\left[\delta^{}_\eta(H-\mu)\right]^{}_{r0},
\end{equation}
where $H$ denotes the Hamiltonian of the system under consideration, or by the density-density correlation function as
\begin{equation}
\label{eq:dd_cond}
\bar\sigma^{}_{} = -\frac{\omega^2}{2} {\rm Tr}\sum_r~r^2_n~G^{}_{0r}\left(\frac{\omega}{2}-\mu+i\eta\right)
G^{}_{r0}\left(\frac{\omega}{2}-\mu-i\eta\right)
\ ,
\end{equation}
where the trace refers to spinor degrees of freedom and $\omega$ to the frequency of an external electric field. The scattering by static disorder has been included here by the phenomenological scattering rate $\eta$. It should be noticed that the prefactor $\omega^2$ in the 
density--density form can be replaced for $\omega\ll\eta$ by $-4\eta^2$, since the transport is controlled by the scattering, represented 
by the scattering rate $\eta$, rather than by the frequency $\omega$ . 
This can be justified by the scaling relation~[\onlinecite{Ziegler2009}]
\begin{eqnarray}
\nn
&\displaystyle
\lim_{\epsilon\to0} {\rm Tr}
\sum_rr_n^2\langle G_{0r}\left(\frac{\omega}{2}+i\epsilon\right) 
G_{r0}\left(-\frac{\omega}{2}-i\epsilon\right)\rangle = 
&\\
\nn
&\displaystyle
\frac{(\omega+2i\eta)^2}{\omega^2}{\omega^2}{\rm Tr}\sum_rr_n^2 G_{0,0r}\left(\frac{\omega}{2}+i\eta\right)
G_{0,r0}\left(-\frac{\omega}{2}-i\eta\right)
\ ,\\
\label{eq:scaling}
\end{eqnarray}
where $G^{}_{0,xy}$ ($G^{}_{xy}$) denote the Green's function of the clean (disordered) system and $\langle\cdot\cdot\rangle$ is 
the averaging with respect to disorder. 
The prefactor $(1+2i\eta/\omega)^2$ appears as a consequence of the integration over the spontaneously broken chiral symmetry. Since $\omega$ breaks the chiral symmetry, the prefactor is finite for $\omega>0$ and diverges only in the dc limit. 
From the point of view of spontaneous symmetry breaking, $\omega$ plays the role of a symmetry breaking field, like the magnetic field
in a ferromagnet, and $\eta$ plays the role of the order parameter (magnetization).   
Replacing this prefactor by 1 gives us the self--consistent Born approximation. This implies that the self-consistent 
Born approximation is valid for high frequencies $\omega\gg\eta$ but not in the dc limit $\omega\ll\eta$. 
In general, the density--density formulation Eq.~(\ref{eq:dd_cond}) can be linked to diffusion via the Einstein relation
~[\onlinecite{Wegner1979a,McKane1981}] (cf. Sect. \ref{sect:discussion}). This is a more direct connection between 
diffusive quantum transport and correlations of the Green's functions than the linear response theory. Moreover, the connection between the two expressions in Eqs.~(\ref{eq:Kubo1}) and (\ref{eq:dd_cond}) is due to the continuity equation 
(cf. Ref.~[\onlinecite{Kubo}]) $\omega\rho+\nabla \cdot {\bf j}=0$, which can be used to replace the current operator by the density operator.
In other words, we use the relation
\[
j_n^2=e^2v_n^2=(e r_n\omega)^2 
\]
in the current--current expression of the conductivity. This was discussed in more detail in Ref.~[\onlinecite{Ziegler2008}].
However, the replacement must be taken with care due to certain limits. In particular, the dc limit $\omega\to 0$
can cause some problems, as we shall discuss in the following. 

The paper is organized as follows: we start with the analysis of the dc conductivity of the clean system 
in Section~\ref{sec:ZerothOrder}. We show that both formulations lead to the same result which is in good 
agreement with the experimentally observed V--like shape of the conductivity as function of the charge density. 
We discuss contributions from different band scattering processes to the total conductivity. While contributions 
from intraband scattering dominate the contribution from interband scattering, the latter are not negligible even at 
large chemical potentials, in contrast to the naive expectation. We point out the subtlety of the reduced current--current 
formulation, i.e. hidden logarithmic divergences arising from each band scattering process. In Section~\ref{sec:PTatZeroCP} 
we proceed with the evaluation of perturbative diagrams of second order in the effective disorder strength at half filling. 
We discuss the effect of different disorder types on the dc conductivity. In current--current language we identify diagrams, 
responsible for logarithmic divergences for a random chemical potential and a random mass. For the case of a random vector potential 
we find in the current--current language finite second order corrections, which is in clear contradiction to the established 
understanding of the physics of this particular disorder acquired from the density--density formula. Generally, we observe an 
entirely different behavior of topologically equivalent diagrams in both formulations, which originates from their different analytical structure. 
In Section~\ref{sec:PTatFinCP} we then compare first order perturbative corrections obtained from both formulas away from the Dirac point. 
We conclude the paper with a discussion and several appendices containing important technical details.

\section{Conductivity at nonzero chemical potential}
\label{sec:ZerothOrder}

Our first goal is to evaluate the dc conductivity of the doped Dirac electron gas within the current--current formulation of Eq. (\ref{eq:Kubo1}).
The $\delta$--functions in this expressions are supposed to have a finite peak width $\eta$, i.e. the scattering rate, which plays the role of the finite ultraviolet cutoff parameter. For Dirac Hamiltonian $H~=~i\nabla\cdot\sigma$ the current operator $j^{}_n$ becomes
\begin{equation}
j^{}_n = -i[H,r^{}_n] = \sigma^{}_n.
\end{equation}
Using the relation between the $\delta$--function and resolvent of an operator $\cal O$
\begin{equation}
\label{eq:delta} 
{\rm Im}[{ \cal O}\pm i\eta]^{-1} = \mp \delta^{}_\eta(\cal O),
\end{equation}
we may replace $\delta$--functions in Eq.~(\ref{eq:Kubo1}) by 
\begin{equation}
[\delta^{}_\eta(H-\mu)]^{}_{xy} = \frac{G^{}_{xy}(-\mu-i\eta) - G^{}_{xy}(-\mu+i\eta)}{2i},
\end{equation}
where $G^{}_{xy}(-\mu \pm i\eta)$ denotes the Green's function of a particle propagating from the spatial point with coordinates $x$ to the point with  coordinates $y$ forwardly 
(if $+i\eta$, i.e. retarded Green's function) or backwardly (if $-i\eta$, i.e. advanced Green's function) in time. This yields the following expression for the conductivity:
\begin{eqnarray}
\nn
\bar\sigma^{}_{} &=& -\frac{1}{2}{\rm Tr}\sum^{}_r~\sigma^{}_n \left[G^{}_{0r}(-\mu-i\eta) - G^{}_{0r}(-\mu+i\eta)\right] \\
\label{eq:Kubo2} 
&\times& \sigma^{}_n \left[G^{}_{r0}(-\mu-i\eta) - G^{}_{r0}(-\mu+i\eta)\right],
\end{eqnarray}
which has been used as the departure point for numerous investigation of transport, mainly in the context of the weak--localization 
approach~[\onlinecite{Ando2002,Suzuura2002,Khveshchenko2006,McCann2006}]. 
In order to distinguish between the different terms we shall call combinations of advanced and retarded Green's functions  
\begin{subequations}
\begin{equation} 
\label{eq:Norm} 
\sim {\rm Tr}\sum_r \sigma^{}_nG^{}_{0r}(-\mu-i\eta)\sigma^{}_nG^{}_{r0}(-\mu+i\eta),
\end{equation}
normal channel, and combinations with only advanced or only retarded Green's functions, e.g.
\begin{equation} 
\label{eq:Anom} 
\sim {\rm Tr}\sum_r \sigma^{}_nG^{}_{0r}(-\mu\pm i\eta)\sigma^{}_nG^{}_{r0}(-\mu\pm i\eta) ,
\end{equation}
\end{subequations}
anomalous channel. Moreover, we employ the Fourier representation to calculate the conductivity. The Green's functions then become 
\begin{equation}
\label{eq:Green} 
G^{}_p (\pm i\eta - \mu) = \frac{\displaystyle \slashed p \mp i(\eta \pm i\mu)}{p^2 + (\eta \pm i\mu)^2},
\end{equation}
where we use the slashed notation $\slashed p = p\cdot \sigma = p_i\sigma_i$. 
Since $\sigma^{}_n\slashed p=p_n\sigma^{}_n\sigma^{}_n-p^{}_{l\neq n}\sigma^{}_l\sigma^{}_n=\slashed p^\dag\sigma^{}_n$, and $\slashed p^\dag \slashed p\sim \cos(2\phi)$ 
vanishes under angular integration, all terms in the numerator containing $p$ vanish all together. After performing the trace with respect to the Dirac matrices we obtain
\begin{eqnarray}
\nn
\bar\sigma^{}_{} = \int_p~\left[\frac{\eta - i\mu}{p^2+(\eta-i\mu)^2} + \frac{\eta+i\mu}{p^2+(\eta+i\mu)^2}\right]^2,
\end{eqnarray}
where $\int_p$ stands for the two--dimensional momentum integration $\int^{+\infty}_{-\infty}d^2p/(2\pi)^2$. For nonzero $\eta$ and $\mu$, the poles in this expression 
lie in the complex plane and we may perform integration along the real axis without going into the complex plane. While
$$
\int_p~\frac{(\eta \pm i\mu)^2}{[p^2+(\eta\pm i\mu)^2]^2} = \frac{\bar\sigma^{}_0}{4},
$$
where $\bar\sigma^{}_0=1/\pi$, another contribution gives 
\begin{eqnarray}
\nn
&\displaystyle\int^{}_p~\frac{2(\eta-i\mu)(\eta+i\mu)}{[p^2+(\eta-i\mu)^2][p^2+(\eta+i\mu)^2]} & \\
\nn
&\displaystyle =\frac{\bar\sigma^{}_0}{2}\frac{\eta^2+\mu^2}{\mu\eta}{\rm atan}\left(\frac{\mu}{\eta}\right).&
\end{eqnarray}
Introducing $z=\mu/\eta$ we obtain for the conductivity
\begin{equation}
\label{eq:Cond} 
\bar\sigma^{}_{} = \frac{\bar\sigma^{}_0}{2}\left(1 + \frac{1+z^2}{z}{\rm atan}(z) \right),
\end{equation}
which is also obtained for the density--density formula in Section~\ref{sec:BandsDD}. This conductivity is quite general for two-dimensional two-band systems with a spectral node \cite{ziegler12}.
At the Dirac point, i.e. for $\mu=0$, this gives the universal minimal conductivity $\bar\sigma^{}_0=1/\pi$ independently of the value of $\eta$. 

\begin{figure}[t]
\includegraphics[height=5cm]{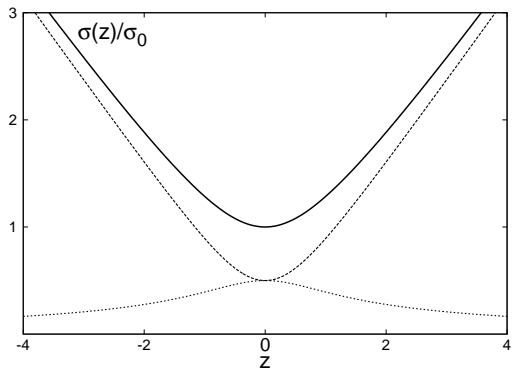} 
\caption{Conductivity formula Eq.~(\ref{eq:Cond}) (solid line) yields the well known dependence on the chemical potential (i.e. the V--shape with respect to charge density)
reported in Ref.~[\onlinecite{Novoselov2005}]. Dotted line shows the conductivity contribution from the interband scattering given in Eq.~(\ref{eq:InterContr}), while the dashed line shows the contribution from the intraband scattering~Eq.~(\ref{eq:IntraContr}).}
\label{fig:Vcond}
\end{figure}

Intuitively, one would expect that only states located in the vicinity of the Fermi--energy $E^{}_F = \mu$ contribute to the conductivity. 
In order to check this statement we calculate explicitly contributions to the conductivity resulting from each band. 
Technical details are given in Appendix~\ref{sec:BandsCC}. Denoting the upper ('conductance') band as the $+$--band and lower ('valence') band as the $-$--band, 
the corresponding contributions from interband scattering are found to be 
\begin{subequations}
\begin{equation}
\label{eq:InterContr} 
\bar\sigma^{}_{+-} = \bar\sigma^{}_{-+} = \frac{\bar\sigma^{}_0}{4z}~{\rm atan}(z)
\ .
\end{equation}
Thus, contributions from interband scattering {\it decrease} with increasing chemical potential $\mu$ or with decreasing scattering rate $\eta$.
On the other hand, contributions from intraband scattering processes are given by
\begin{equation}
\label{eq:IntraContr} 
\bar\sigma^{}_{++} = \bar\sigma^{}_{--} = \frac{\bar\sigma^{}_0}{4}(1+z~{\rm atan}(z)),
\end{equation}
\end{subequations}
which {\it increase} with increasing chemical potential or decreasing scattering rate (cf. Fig.~\ref{fig:Vcond}). 

The reduced current--current formula, which takes only the normal channel of Eq.~(\ref{eq:Norm}) into account, is sometimes used for the conductivity calculations~[\onlinecite{Ando2002,Suzuura2002,Khveshchenko2006,McCann2006}]. For this we consider as previously the projections onto the bands and obtain 
\begin{subequations}
\begin{equation} 
\label{eq:PppNorm}
\bar\sigma^{N}_{++} = \bar\sigma^{N}_{--} =  \frac{\bar\sigma^{}_0}{4}(\ell + z~{\rm atan}(z)) , \ \ \ 
\ell=\frac{1}{2}\log\left(\frac{\Lambda^2}{\eta^2+\mu^2}\right)
\end{equation}
for the intraband and 
\begin{equation} 
\bar\sigma^{N}_{+-} = \bar\sigma^{N}_{-+} =  \frac{\bar\sigma^{}_0}{4}\left(-\ell + \frac{1}{z}~{\rm atan}(z)\right)
\end{equation}
\end{subequations}
for the interband contributions. Here we have introduced the momentum cutoff $\Lambda$. While Eq.~(\ref{eq:Norm}) in total is finite, each projection acquires a logarithm, with different sign though, which may become very large at the Dirac point ($\mu\to 0$) due to the smallness of $\eta$. Eq.~(\ref{eq:PppNorm}) is valid only for arguments of the logarithms of order unity, i.e. for the 
$\mu\sim\Lambda$($\sim 1$eV for graphene). This has led many authors to the conclusion that the perturbative calculation is only
valid far away from the Dirac point (cf. \cite{Suzuura2002}). In contrast to this statement, however, the inclusion of both bands in the conductivity leads to the cancellation of the logarithmic terms and we obtain a finite conductivity even at the Dirac node for arbitrarily weak scattering.

\section{Second order perturbation theory at half filling}
\label{sec:PTatZeroCP}

\begin{figure}[t]
\includegraphics[height=1.2cm]{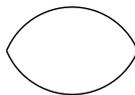}
\caption{Zeroth order diagram}
\label{fig:0thOrd}
\end{figure}
\vspace{5mm}
\begin{figure}[t]
\includegraphics[width=6.5cm]{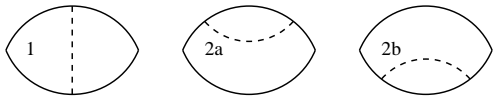}
\caption{First order diagrams}
\label{fig:1stOrd}
\end{figure}
\vspace{5mm}
\begin{figure}[h]
\includegraphics[width=6.5cm]{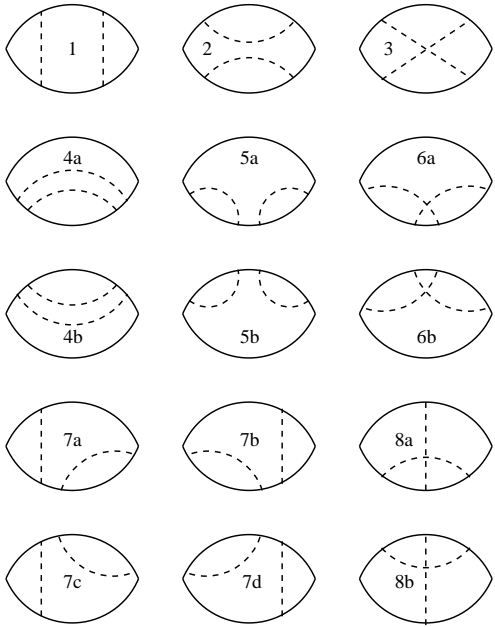}
\caption{Second order diagrams}
\label{fig:2ndOrd}
\end{figure}

While both expressions for the conductivity in Eqs. (\ref{eq:Kubo1}), (\ref{eq:dd_cond}) give the same result in the self--consistent approximation, it would be interesting to see whether this also holds in perturbation theory with respect to disorder. 
The expressions have very different analytical structures and, therefore, they may behave differently once disorder is included. The envisaged task aims at the general understanding of this behavior and shall be pursued at the simplest possible level. Therefore, we study the conductivity perturbatively, assuming disorder to be weak. In our previous paper [\onlinecite{Sinner2011}] we already studied the conductivity in the density--density formula  and came to the conclusion that the perturbation series is free of logarithmic divergences, at least up to second order. Here we perform an analogous investigation in the current--current formula. The relative simplicity of this formulation in comparison to the density--density language enables us to perform all calculations analytically in a fully controllable way. We consider in detail the case of the random scalar potential which plays the role of the spatially fluctuating chemical potential. 
We start with Eq.~(\ref{eq:Kubo2})
\begin{eqnarray}
\nn
\langle\bar\sigma^{}_{}\rangle &=& -\frac{1}{2}{\rm Tr}\sum^{}_r~\langle\sigma^{}_n \left[G(-v-i\eta) - G(-v+i\eta)\right]^{}_{0r} \\
\label{eq:Kubo4} 
&\times& \sigma^{}_n \left[G(-v-i\eta) - G(-v+i\eta)\right]^{}_{r0}\rangle,
\end{eqnarray}
where  $\langle\cdot\cdot\rangle$ denotes averaging with respect to a random chemical potential $v$ with mean zero and with the Gaussian correlator
\begin{equation}
\langle v^{}_r v^{}_{r^\prime}\rangle = g\delta(r-r^\prime). 
\end{equation}
It is convenient to consider both normal and anomalous channels separately and sum over all contributions after the perturbative calculations are performed. 
Obviously, this is allowed by the linearity of the averaging process (a usual Gaussian integration). We emphasize that it is crucial to account for both, normal and anomalous channels. 

Taking the degeneracy into account, the conductivity in the normal channel reads 
\begin{equation}
\label{eq:Normal}
\langle\bar\sigma^{}_{N}\rangle = {\rm Tr}\sum^{}_r~\langle\sigma^{}_n G^{}_{0r}(-v-i\eta)\sigma^{}_n G^{}_{r0}(-v+i\eta)\rangle, 
\end{equation}
while in the anomalous channel we have
\begin{equation}
\label{eq:Anomal}
\langle\bar\sigma^{}_{A}\rangle = -{\rm Tr}\sum^{}_r~\langle\sigma^{}_n G^{}_{0r}(-v-i\eta)\sigma^{}_n G^{}_{r0}(-v-i\eta)\rangle. 
\end{equation}
The topology of the diagrams is the same in both channels. Fig.~\ref{fig:0thOrd} shows the zeroth order diagram, contributing to the universal dc conductivity. 
There are three diagrams in each channel in first order, as depicted in Fig.~\ref{fig:1stOrd}  and fifteen diagrams in second order shown in Fig.~\ref{fig:2ndOrd}. 

\begin{widetext}

\begin{table}[h]
\begin{tabular}{ccccc}
  \hspace{2mm}   Diagram  \hspace{2mm}    &  \hspace{2mm}  {\rm Normal channel} \hspace{2mm}  & \hspace{2mm}  {\rm Anomalous channel} \hspace{2mm} & \hspace{2mm} {\rm Both channels} \hspace{2mm} \hspace{2mm} & {\rm Ref.~[\onlinecite{Sinner2011}]}\hspace{2mm}   \\
\\
\includegraphics[height=3mm]{fig2.eps}  $\times$ 1
&  
1/2
& 
1/2
&
1
&
1
\end{tabular}
\caption{Contributions to universal conductivity in both channels from the zeroth--order diagram Fig.~\ref{fig:0thOrd} in units of $\bar\sigma^{}_0$.}
\label{tab:0thOrd} 
\vspace{5mm}
\begin{tabular}{ccccc}
  \hspace{2mm}   Diagram  \hspace{2mm}    &  \hspace{2mm}  {\rm Normal channel} \hspace{2mm}  & \hspace{2mm}  {\rm Anomalous channel} \hspace{2mm} & \hspace{2mm} {\rm Both channels} \hspace{2mm} \hspace{2mm} & {\rm Ref.~[\onlinecite{Sinner2011}]}\hspace{2mm}   \\
\\
\includegraphics[height=3mm]{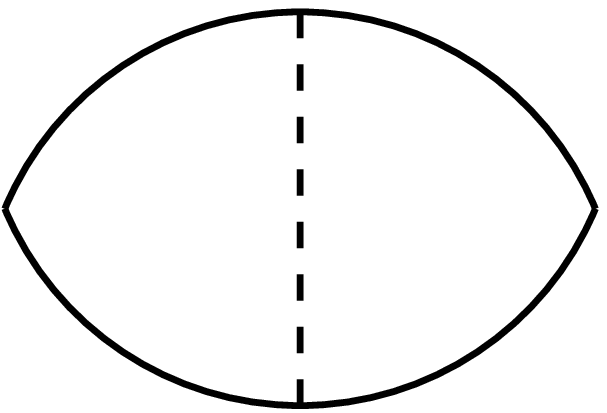}  $\times$ 1
&  
$\alpha/4$
& 
$-\alpha/4$
&
0
&
$\alpha[1+2\ell]$

\\
\\
\includegraphics[height=3mm]{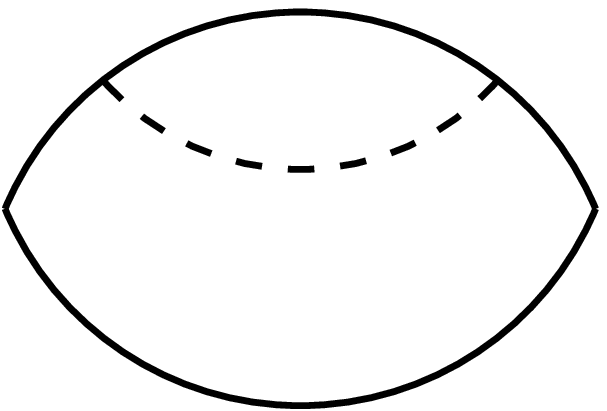} $\times$ 2 
& 
0
&
0
& 
0
&
$\displaystyle-2\alpha\ell$
\\
\\
{\rm Total} 
& 
$\alpha/4$
&
$-\alpha/4$
&
0
&
$\alpha$  
\end{tabular}
\caption{Conductivity corrections from the first--order diagrams Fig.~\ref{fig:1stOrd} in units of the universal conductivity $\bar\sigma^{}_0$, $\alpha=g/2\pi$, $\ell=\log\Lambda/\eta$.}
\label{tab:1stOrd} 
\vspace{5mm}
\begin{tabular}{ccccc}
  \hspace{2mm}   Diagram   \hspace{2mm}  &  \hspace{2mm} {\rm Normal channel} \hspace{2mm}  &  \hspace{2mm}  {\rm Anomalous channel} \hspace{2mm} &\hspace{2mm}  {\rm Both channels} \hspace{2mm}  & \hspace{2mm}  {\rm Ref.~[\onlinecite{Sinner2011}]}\hspace{2mm} \\
\\
\includegraphics[height=3mm]{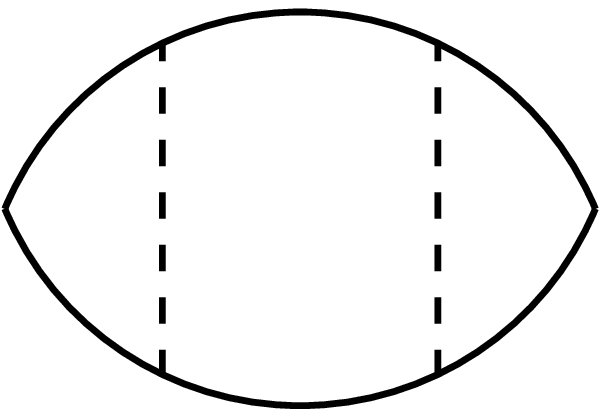} $\times$ 1 
&  
$\alpha^2/8$
& 
$\alpha^2/8$
&
$\alpha^2/4$
&
${\alpha^2}[1 + 4\ell + 6\ell^2]/2$
\\
\\
\includegraphics[height=3mm]{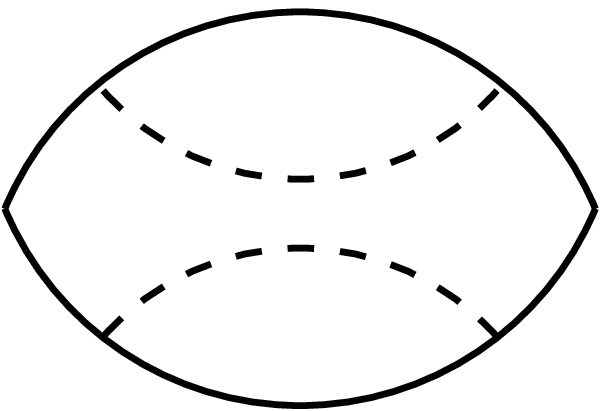} $\times$ 1
& 
$\alpha^2\ell^2/6$
&
$\alpha^2\ell^2/6$
& 
$\alpha^2\ell^2/3$
&
$5\alpha^2\ell^2/3$
\\
\\
\includegraphics[height=3mm]{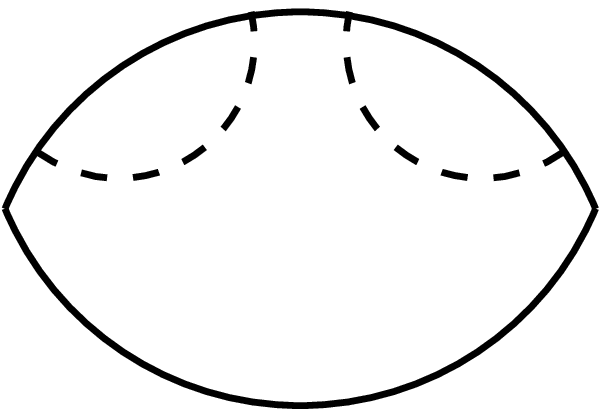} $\times$ 2
&
$-\alpha^2\ell^2/6$
&
$-\alpha^2\ell^2/6$
& 
$-\alpha^2\ell^2/3$
&
$4\alpha^2\ell^2/3$
\\
\\
\includegraphics[height=3mm]{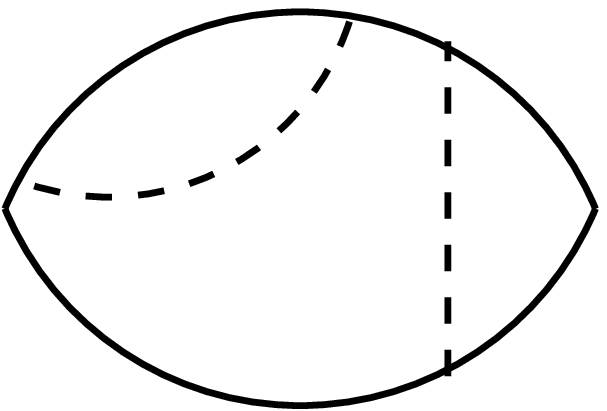} $\times$ 4
&
0
&
0
&
0
&
$-4\alpha^2\ell[1+\ell]$
\\
\\
\includegraphics[height=3mm]{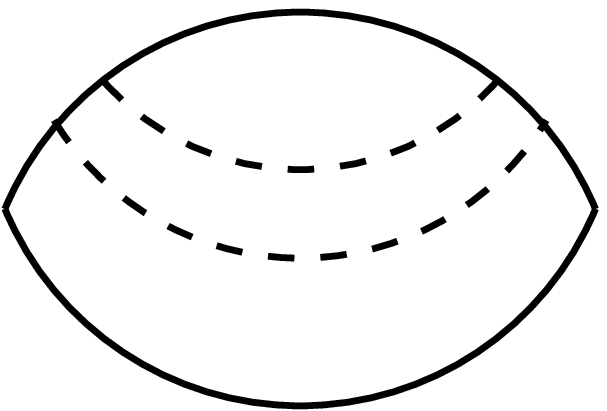} $\times$ 2
&
0
&
0
&
0
&
$2\alpha^2\ell[1-\ell]$
\\
\\
\includegraphics[height=3mm]{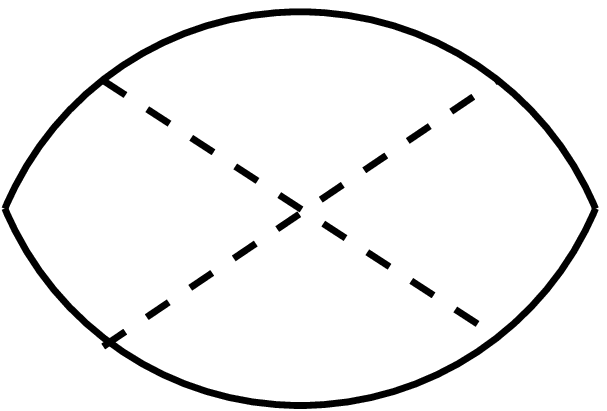} $\times$ 1 
& 
$-\alpha^2[c - 2\ell]/4$
& 
$-\alpha^2[c - 2\ell]/4$
& 
$-\alpha^2[c - 2\ell]/2$
&
$ 0 $ 
\\
\\
\includegraphics[height=3mm]{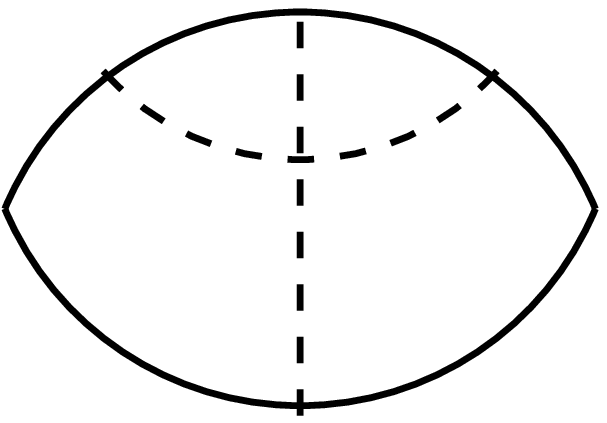} $\times$ 2
&
$-\alpha^2[c-2\ell]/2$
& 
$\alpha^2[c - 2\ell]/2$
& 
$ 0 $
&
$ 0 $ 
\\
\\
 \includegraphics[height=3mm]{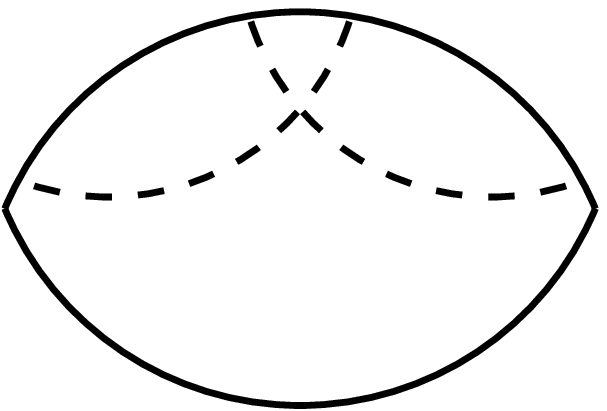} $\times$ 2 
& 
$-\alpha^2[c-2\ell]/4$
& 
$-\alpha^2[c-2\ell]/4$
& 
$-\alpha^2[c-2\ell]/2$
& $ 0 $ 
\\
\\
{\rm Total} 
& 
$\alpha^2 (1 - 8 c + 16\ell)/8$
&
$\alpha^2/8$
& 
$\alpha^2(1 - 4 c + 8\ell)/4$
& $\displaystyle  {\alpha^2}/{2}$  
\end{tabular}
\caption{Conductivity corrections from the second--order diagrams Fig.~\ref{fig:2ndOrd} in units of the universal conductivity $\bar\sigma^{}_0$, $\alpha=g/2\pi$, $\ell=\log\Lambda/\eta$. 
The constant $c$ is calculated in Eq.~(\ref{eq:cdef}) with the numerical value $\sim 3.1036$.}
\label{tab:2ndOrd} 
\end{table}

\end{widetext}

The detailed evaluation of all first and second order diagrams in the current--current formula is represented in Appendix~\ref{sec:PertCC}. 
The evaluation of first order diagrams in the density--density formula is shown in Appendix~\ref{sec:BandsDD}. The results of those calculations 
are summarized in Tables~\ref{tab:1stOrd} and \ref{tab:2ndOrd}. The difference between the two formulas could not be more striking. Virtually none 
of the diagrams have the same value in both formulas. Both normal and anomalous channels in the current--current formula do not 
contribute to the conductivity on the same footing but seem to compete at every order. As one of the results of this competition, there is 
no conductivity correction to order $g$ at all, while in the density--density formula such a correction certainly exists. A similar situation 
is also observed to the order $g^2$. In analogy to the density--density formula we see here that diagrams with and without intersection of 
impurity lines, i.e. $\includegraphics[height=3mm]{fig7.eps}$, $\includegraphics[height=3mm]{fig9.eps}$, both $\includegraphics[height=3mm]{fig11.eps}$, 
and all four $\includegraphics[height=3mm]{fig14.eps}$ on one hand and $\includegraphics[height=3mm]{fig10.eps}$, both $\includegraphics[height=3mm]{fig13.eps}$, 
and both $\includegraphics[height=3mm]{fig15.eps}$ on the other, build up disjoint subsets. Logarithmic divergences of diagrams without disorder line intersections 
cancel exactly in both formulas. This is even the case in both channels in the current--current formula, in an entirely different way though. On the contrary, 
diagrams with intersecting disorder lines behave differently not only in both formulas, but also in both channels in the current--current formula. 
In density--density formula the structure is too complicated to be evaluated analytically. Their evaluation with Mathematica in Ref.~[\onlinecite{Sinner2011}] 
led us to the conclusion that each of them vanish separately by angular integration. In contrast, the analytical structure of these diagrams in the current--current 
language makes an analytical evaluation possible. It turns out that each of these diagrams diverges only logarithmically, in contrast to diagrams without
intersections, which diverge as a squared logarithm. The amplitudes of the topologically equivalent diagrams are the same in both channels but the sign is 
different such that these logarithms cancel each other in the anomalous channel but sum themselves up in the normal channel to a global logarithmic singularity. 
Taking all the contributions to order $g^2$, the conductivity in current--current formula reads
\begin{equation}
\label{eq:CorrCond} 
\frac{\langle\bar\sigma\rangle}{\bar\sigma^{}_0} \approx 1 + \frac{\alpha^2}{4} + 2\alpha^2(\ell-c^{}_v),
\end{equation}
where $\alpha=g/2\pi$, $c^{}_v\sim1.5518$ and $\ell=\log\Lambda/\eta$. The logarithmic divergence in Eq.~(\ref{eq:CorrCond}) 
looks very similar to the usual weak--antilocalization term~[\onlinecite{Hikami1980}], but it is not, since the latter 
is due to the massless cooperon mode which cannot be seen in any finite order of the perturbative expansion. This divergence resembles more
what was called the {\it ultraviolet logarithmic corrections} in Ref.~[\onlinecite{Aleiner2006}]. 

In analogy to the case of random chemical potential we are able to calculate perturbative 
corrections also for other disorder types. Below we briefly summarize results of these calculations. 

{\it Random mass:}  The case of the random mass is of particular interest, since it governs a metal--insulator transition. 
This disorder type has been intensively studied for the couple of decades, both analytically~[\onlinecite{Ziegler1997,Ziegler2009,Bocquet2000}] and more recently 
numerically~[\onlinecite{Bardarson2010,Medvedeva2010}]. This potential couples to the Pauli matrix $\sigma^{}_3$ and anticommutes with the Hamiltonian of the clean system. 
This leads to some differences 
in the calculation. Again we see significant discrepancies with our previous work Ref.~[\onlinecite{Sinner2011}] in both orders. All diagrams which vanish for the random chemical potential 
do vanish for the random mass as well, in particular diagrams $\includegraphics[height=3mm]{fig8.eps}$, $\includegraphics[height=3mm]{fig11.eps}$ 
and $\includegraphics[height=3mm]{fig14.eps}$. Contributions from diagrams $\includegraphics[height=3mm]{fig6.eps}$ and $\includegraphics[height=3mm]{fig15.eps}$ 
average to zero after summing over both channels. As in the case of the random scalar potential we also observe 
$\includegraphics[height=3mm]{fig9.eps}+2~\includegraphics[height=3mm]{fig12.eps}=0$ in both channels independently.  
Therefore, the leading order correction is again of the order $g^2$. Diagram $\includegraphics[height=3mm]{fig7.eps}$ is finite while diagrams $\includegraphics[height=3mm]{fig10.eps}$ and $\includegraphics[height=3mm]{fig13.eps}$ diverge logarithmically. The total conductivity to second order reads
\begin{equation}
\frac{\langle\bar\sigma\rangle}{\bar\sigma^{}_0} \approx 1 + \frac{\alpha^2}{4} + 2\alpha^2(\ell+c^{}_m)
\end{equation}
with the constant $c^{}_m\approx 1.1052$. Interestingly, the logarithm was not observed numerically in Ref.~[\onlinecite{Bardarson2010}].

{\it Random vector potential:} This disorder type is usualy associated with the surface corrugations  which arise due to thermal instability of two--dimensional crystals. It was shown in the
density--density formalism~[\onlinecite{Ludwig1994,Bhaseen2000}] that this disorder type does not change the universal value of the conductivity, so we might expect this to be true also in
the current--current formula. Because of the vector nature of this potential $v\to a^{}_i\sigma^{}_i$, $i=1,2$ the Gaussian correlator is changed to 
\begin{equation}
\langle a^{}_i(x)a ^{}_j(x^\prime)\rangle = g\delta^{}_{ij} \delta(x-x^\prime).
\end{equation}
All diagrams which give zero contributions for the other disorder types vanish for the case of the random vector potential too. Also the relation $\includegraphics[height=3mm]{fig9.eps}+2~\includegraphics[height=3mm]{fig12.eps}=0$ holds in both channels. The major difference to other disorder types consist in the behavior of the diagrams with intersecting disorder lines: they do not develop any logarithmic divergences. Instead, they obey a beautiful relation 
$\includegraphics[height=3mm]{fig10.eps}+2~\includegraphics[height=3mm]{fig13.eps} + 2~\includegraphics[height=3mm]{fig15.eps} =0$ in each channel with the relative weights of each diagram respectively $1, 1/2, -1$. However, diagram $\includegraphics[height=3mm]{fig7.eps}$ yields a finite contribution which is not compensated by any other terms. In contrast to the exact result of Ref.~[\onlinecite{Ludwig1994}] we find  that the conductivity acquires finite corrections as
\begin{equation} 
\frac{\langle\bar\sigma\rangle}{\bar\sigma^{}_0} \approx 1 + \alpha^2,
\end{equation}
which clearly deviates from the previous results.

\section{Leading order perturbation theory at nonzero chemical potential}
\label{sec:PTatFinCP}

Our next task is to extend the perturbative analysis to a nonzero chemical potential. In this section we compute the leading order perturbative 
corrections in disorder strength to Eq.~(\ref{eq:Cond}). 
We start with Eq.~(\ref{eq:Kubo2}), average over disorder and reorder terms due to each channel:
\begin{subequations}
\begin{eqnarray}
\label{eq:AvKubo2_1} 
&\displaystyle
\langle\bar\sigma^{}_{}\rangle = {\rm Tr}\int^{}_p~\langle\sigma^{}_n G^{}_p(-\mu-i\eta) \sigma^{}_n G^{}_p(-\mu+i\eta)\rangle 
&\\
\label{eq:AvKubo2_2} 
&\displaystyle
- \frac{1}{2}{\rm Tr}\sum_{s=\pm}\int^{}_p~\langle\sigma^{}_n G^{}_p(-\mu+is\eta) \sigma^{}_n G^{}_p(-\mu+is\eta)\rangle,&
\hspace{4mm}
\end{eqnarray}
\end{subequations}
Expanding to leading order in $g$ we obtain from Eq.~(\ref{eq:AvKubo2_1}) 
\begin{subequations}
\begin{eqnarray} 
\nn
&
\displaystyle g {\rm Tr} \int^{}_p~G^{}_p(-\mu-i\eta)\sigma^{}_n G^{}_p(-\mu+i\eta) 
&\\
\label{eq:AvNCh1}
&\times
\displaystyle \int^{}_q~G^{}_q(-\mu+i\eta)\sigma^{}_n G^{}_q(-\mu-i\eta)
&\\
\nn
&
\displaystyle + g{\rm Tr}\sum_{s=\pm}\int^{}_q~G^{}_q(-\mu+is\eta)\times
&\\
\nn
&
\displaystyle  \int^{}_p G^{}_p(-\mu+is\eta)\sigma^{}_nG^{}_p(-\mu-is\eta)\sigma^{}_nG^{}_p(-\mu+is\eta),
&\\
\label{eq:AvNCh2}
\end{eqnarray}
and from Eq.~(\ref{eq:AvKubo2_2}) 
\begin{eqnarray}
\nn
&
\displaystyle -\frac{g}{2}{\rm Tr}\sum_{s=\pm}\int_p G^{}_p(-\mu+is\eta)\sigma^{}_nG^{}_p(-\mu+is\eta) 
& \\
\label{eq:AvACh1}
&
\displaystyle\times\int^{}_q G^{}_q(-\mu+is\eta)\sigma^{}_n G^{}_q (-\mu + is\eta)
&\\
\nn
&
\displaystyle -g {\rm Tr}\sum_{s=\pm}\int_q G^{}_q(-\mu+is\eta)\times
&\\
\nn
&
\displaystyle
\int_p
G^{}_p(-\mu +is\eta)\sigma^{}_nG^{}_p(-\mu +is\eta)\sigma^{}_nG^{}_p(-\mu +is\eta).
&\\
\label{eq:AvACh2}
\end{eqnarray}
\end{subequations}
Diagrammatically, Eqs.~(\ref{eq:AvNCh1}) and (\ref{eq:AvACh1}) correspond to $\includegraphics[height=3mm]{fig6.eps}$ and Eqs.~(\ref{eq:AvNCh2}) and (\ref{eq:AvACh2}) to both diagrams $\includegraphics[height=3mm]{fig8.eps}$ which become different (complex conjugated of each other) at nonzero chemical potential. Eqs.~(\ref{eq:AvACh1}) and (\ref{eq:AvACh2}) represent contributions to the conductivity from the anomalous channel and are much easier to evaluate. Because they contain either only retarded or only advanced Green's functions, the evaluation does not differ much from the undoped case. Eq.~(\ref{eq:AvACh2}) gives zero and 
\begin{equation}
{\rm Eq.~(\ref{eq:AvACh1})} = -\frac{\bar\sigma^{}_0}{4}\alpha.
\end{equation}
In contrast, Eqs.~(\ref{eq:AvNCh1}) and (\ref{eq:AvNCh2}) require more effort. Concerning Eq.~(\ref{eq:AvNCh2}),
we obtain after performing $q$--integral, the trace and angular integral in $p$--part
\begin{eqnarray}
\nn
{\rm Eq.~(\ref{eq:AvNCh2})} &=& 2g\sum_{s=\pm}\frac{\eta^2+\mu^2}{(4\pi)^2}\log\left(\frac{\Lambda^2}{\eta^2+\mu^2}e^{-2i s {\rm atan}\frac{\mu}{\eta}} \right)\\
\nn
&\times& \int_0^\infty dt~\frac{t-(\eta+is\mu)^2}{[t+(\eta+is\mu)^2]^2[t+(\eta-is\mu)^2]},
\end{eqnarray}
where $t=p^2$. After performing the partial fraction decomposition under the  $t$--integral 
$$
\frac{t-A^2}{[t+A^2]^2[t+B^2]} = \frac{\alpha}{t+A^2}-\frac{\alpha}{t+B^2}+\frac{\beta}{[t+A^2]^2}
$$
with
$$
\alpha= \frac{A^2+B^2}{(A^2-B^2)^2},\;\;\; \beta = \frac{2A^2}{A^2-B^2},
$$
it can be easily evaluated. We obtain 
\begin{eqnarray}
\nn
&\displaystyle{\rm Eq.~(\ref{eq:AvNCh2})} = \frac{4g}{(2\pi)^2} \sum_{s=\pm}\frac{\eta^2+\mu^2}{4\mu\eta}&\\
\nn
&\times\displaystyle\left[-2 {\rm atan}\frac{\mu}{\eta} - is\log\frac{\Lambda^2}{\eta^2+\mu^2}\right]
\left[1 - \frac{\eta^2-\mu^2}{\eta\mu}{\rm atan}\frac{\mu}{\eta}\right],
&
\end{eqnarray}
such that the imaginary parts cancel each other upon summation over $s$ and the final result does not depend on the cutoff $\Lambda$. 
Eventually we arrive at the following total contribution from all $\includegraphics[height=3mm]{fig8.eps}$--like diagrams:
\begin{equation}
\label{eq:RainBow}
\sum_{\rm all} {\includegraphics[height=3mm]{fig8.eps}} = -\frac{\bar\sigma^{}_0\alpha}{2}\frac{1+z^2}{z}{\rm atan}(z)~
\left(1 - \frac{1-z^2}{z}{\rm atan}(z) \right).
\end{equation}
The remaining contribution Eq.~(\ref{eq:AvNCh1}) corresponds to the normal channel correction from the ${\includegraphics[height=3mm]{fig6.eps}}$--diagram. Since both $q$-- and $p$--integrals give equal contributions to the final result, we obtain
\begin{eqnarray}
\nn
{\rm Eq.~(\ref{eq:AvNCh1})} &=& \frac{\bar\sigma^{}_0\alpha}{4}\left(\int_0^\infty dt~\frac{\eta^2+\mu^2}{[t+(\eta+i\mu)^2][t+(\eta-i\mu)^2]} \right)^2\\
\nn
&=& \frac{\bar\sigma^{}_0\alpha}{4}
\left(
\frac{1+z^2}{z}{\rm atan}(z)
\right)^2.
\end{eqnarray}
Then, the total contribution from all ${\includegraphics[height=3mm]{fig6.eps}}$--like diagrams reads
\begin{equation}
\label{eq:Vertical} 
\sum_{\rm all} {\includegraphics[height=3mm]{fig6.eps}} = \frac{\bar\sigma^{}_0\alpha}{4}\left[ 
\left(\frac{1+z^2}{z}{\rm atan}(z)\right)^2 - 1 \right].
\end{equation}
Eqs.~(\ref{eq:RainBow}) and (\ref{eq:Vertical}) together give the total first order correction in $g$ for Eq.~(\ref{eq:Cond}). 
It vanishes for $z\to0$ in accord to our calculations of Section~\ref{sec:PTatZeroCP}. In the other limit $z\to\infty$ it becomes negative
\begin{equation}
\frac{\langle\bar\sigma\rangle-\bar\sigma}{\bar\sigma^{}_0}\sim -\alpha\frac{\pi^2 z^2}{16},
\end{equation}
where $\bar\sigma$ is given by Eq.~(\ref{eq:Cond}). For comparison, in Appendix~\ref{sec:BandsDD} we have evaluated the 
same diagrams in the density--density formalism with the result given in Eq.~(\ref{eq:1stCorrDD}). While for small $z$, 
both expressions behave very differently, the large $z$ behavior is similar. In both cases it goes $\sim - z^2$ with 
the only difference of factor 2 in the weight. For small $z$ both expressions are plotted in Fig.~\ref{fig:2ndOrd}. 

\begin{figure}[t]
\includegraphics[height=5cm]{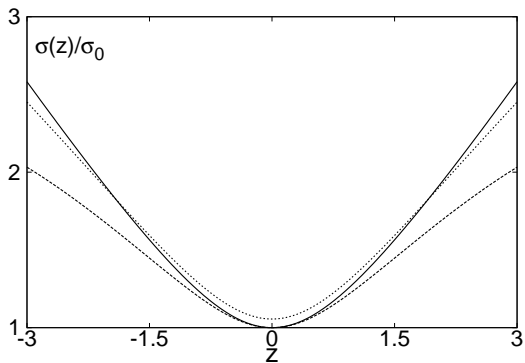} 
\caption{Corrected dc conductivity in comparison to the conductivity in Born--approximation Eq.~(\ref{eq:Cond}) (solid line). 
Dashed curve shows the result from the current--current formula, dotted line that from the density--density formula plotted for $\alpha\approx0.1$.}
\label{fig:Comparison}
\end{figure}

\section{discussion}
\label{sect:discussion}

In this paper we have carried out a comparative analysis of two frequently used particular formulations of the Kubo conductivity formula 
for the case of a weakly disordered two--dimensional Dirac electron gas in the dc limit, the  current--current and density--density formulas. 
Even a superficial glimpse at these formulas, as they are given by Eqs.~(\ref{eq:Kubo1}) and (\ref{eq:dd_cond}), suffices to recognize that
they are structurally different from each other. Mainly, the difference is due to the fact that the current operator is up to a constant a 
non--diagonal Pauli matrix. Therefore, the intermixing of the bands occurs in the current--current formula, whereas in the density--density 
formula this is not the case. Fortunately, however, for the clean and even for the disordered system in self--consistent Born approximation 
both formulas provide us with the very same expression at and away from the Dirac point, as it is given in Eq.~(\ref{eq:Cond}) and shown in Fig.~\ref{fig:Vcond}. 
Generally, we conclude that the dc transport behavior of the two--dimensional Dirac electron gas becomes more conventional 
as we go deeper in the conduction band, because interband scattering becomes less important. The difference of both formulations, however, becomes 
evident if one computes corrections to the conductivity in terms of disorder strength. To leading order they are shown in Fig.~\ref{fig:Comparison} 
as functions of the chemical potential. The correction obtained for the density--density formula reveals a weaker dependence on the chemical potential, 
since it barely deviates from the Dirac point value up to $z=\pm 1$. In contrast, the correction obtained from the current--current formula reveals 
a substantial deviation from the Dirac point value already at $z\sim\pm0.5$. 

The properties of both conductivities do not deviate dramatically in our calculation. A qualitative comparison with the experimentally
observed V shape of the conductivity as a function of the charge density~[\onlinecite{Novoselov2005}] can be matched by both
expressions, after fitting the scattering rate $\eta$ properly (cf. Fig. \ref{fig:Comparison}). Thus, a comparison with experimental data cannot
decide over the validity of the two Kubo formulas, as long as we do not have accurate measurements of the scattering 
rate.  As we go to higher orders of corrections we may find stronger deviations. This is indicated by the results
of the weak-localization approach, which includes partial summations of infinitely many perturbation terms.
The current-current correlation function provides in this case a logarithmically increasing 
conductivity~[\onlinecite{{McCann2006}}] 
\begin{equation}
\sigma=\frac{e^2}{h}\left[\frac{8}{g}+\frac{4}{\pi}\log\left(\frac{\mu}{\mu^{}_\phi}\right)\right]
\ ,
\end{equation}
where $\mu_\phi$ is a phenomenological parameter due to inelastic scattering which vanishes for a vanishing temperature.
In contrast to this expression the density-density correlation function gives us Eq. (\ref{eq:Cond}) again~[\onlinecite{ziegler13}]. 
These results clearly indicate that only the density-density correlation gives the V-shape conductivity.

A more direct support for the density--density formula comes from the Einstein relation: the physics of dc transport in 
the presence of weak scattering should be governed by the diffusion of quasiparticles. This relation has
been used in many studies of disorder scattering to express the conductivity through the density-density
correlation for one-band metals~\cite{Wegner1979a,McKane1981} and for two-band metals~[\onlinecite{ziegler12,Ziegler2012}]. 
For this purpose we introduce the probability for a quasiparticle to move from site $\br'$ to site $\br$ during the time $t$
with probability $P_{\br\br'}(t)=|\langle\br|\exp(-iHt)|\br'\rangle|^2$, where $H$ is the hopping Hamiltonian. 
Then the mean square displacement with respect to $\br'=0$ describes diffusion if the following equation is satisfied:
\begin{equation}
\langle r_k^2\rangle=\sum_\br r_k^2 P_{\br,0}(t)=Dt
\ .
\label{diff2}
\end{equation}
Using the Green's function we obtain for large distances $|\br-\br'|$ and $\epsilon\sim 0$
\begin{equation}
\int_0^\infty P_{\br\br'}(t)e^{-\epsilon t}dt
\sim \int_{E_0}^{E_F}\langle G_{\br\br'}(E+i\epsilon)G_{\br'\br}(E-i\epsilon)\rangle dE
\ ,
\label{gf2}
\end{equation}
where $E_0$ is the lower band edge.
Then we get from Eq. (\ref{gf2}) for the diffusion coefficient at energy $E$
\begin{equation}
D(E)\sim\lim_{\epsilon\to0}\epsilon^2\sum_\br  r_k^2\langle G_{\br0}(E+i\epsilon)G_{0\br}(E-i\epsilon)\rangle_d
\label{diff_c0}
\end{equation}
with $D=\int D(E)\rho(E)dE$ in Eq. (\ref{diff2}) and with the density of states $\rho$. For transport at low
temperatures we need the diffusion coefficient only at the Fermi energy $E_F$. This expression agrees with the 
density--density formula (\ref{eq:dd_cond}), since the dc conductivity can be calculated from $D(E)$ via the Einstein 
relation $\sigma\propto \rho(E_F) D(E_F)$.

In conclusion, we have found that perturbative corrections for both conductivity formulas differ substantially from each other in 
every order and even for every type of diagrams. This is a problem of the dc limit, since the equivalence of the formulas for $\omega\gg \eta$ 
is easy to show \cite{Ziegler2008}. Our findings question the comparability of results acquired by different calculational methods, since in some
cases the density--density formula, in other cases the current--current formula have been used. From our comparison with diffusion we would conclude that the density--density formula is preferable. A challenge would to reformulate the weak--localization approach in terms of the density--density formula and to compare it with the nonlinear sigma--model approach, which is based on the density--density formula. 
{\bf
A first attempt in this direction can be found in Ref.~[\onlinecite{ziegler13}].
}

\section*{ACKNOWLEDGMENTS}

We acknowledge useful discussions with E. Hankiewicz and R. Raimondi. 

\appendix

\section{Contributions to conductivity from different band scattering processes}
\label{sec:BandsCC}

In order to estimate the relative importance of interband versus intraband scattering it is convenient to go into the 
representation in which Green's functions are diagonal. The unitary transformation diagonalizing them reads
\begin{equation}
 U^{}_p = \frac{1}{\sqrt{2}}\left( 
\begin{array}{ccc}
\displaystyle  \frac{p^{}_1 - ip^{}_2}{p} & \displaystyle \frac{p^{}_1 - ip^{}_2}{p} \\
\\
1 & -1 
\end{array}
\right),
\end{equation}
with $p=\sqrt{p^2_1+p^2_2}$. In the diagonal representation Green's functions become $\hat G^{}_p = U^\ast_p G^{}_p U^{}_p$, i.e.
\begin{equation}
\hat G^{}_{p}(\pm i\eta-\mu) = 
\frac{p\sigma^{}_3 \mp i (\eta \pm i\mu)}{p^2 + (\eta\pm i\mu)^2}.
\end{equation}
The conductivity in diagonal representation reads
\begin{eqnarray}
\nn
\bar\sigma^{}_{} &=& \displaystyle -\frac{1}{2}{\rm Tr}\int^{}_p~
\left[
\hat G^{}_p(-i\eta-\mu) - \hat G^{}_p(i\eta-\mu)
\right]\\
\label{eq:Kubo3d}
&\times&
\hat\sigma^{}_n \sigma^{}_0 
\left[
\hat G^{}_p(-i\eta-\mu) - \hat G^{}_p(i\eta-\mu)
\right]
\hat\sigma^{}_n \sigma^{}_0 , 
\end{eqnarray}
where current operators in diagonal representation are 
\begin{equation}
\hat\sigma^{}_n =  U^\ast_p\sigma^{}_n U^{}_p.
\end{equation}

The contributions from each band scattering process can be found by expanding the unity operator $ \sigma^{}_0 $ in basis of orthogonal projectors on each subband:
\begin{equation}
\sigma^{}_0 = {\cal P}^{}_+ + {\cal P}^{}_- = \frac{\sigma^{}_0 +  \sigma^{}_3}{2} + \frac{\sigma^{}_0 - \sigma^{}_3}{2}.
\end{equation}
The projectors act on the Green's functions as follows
\begin{equation}
{\cal P}^{}_\pm \frac{p\sigma^{}_3 \mp i (\eta \pm i\mu)}{p^2 + (\eta\pm i\mu)^2} = {\cal P}^{}_\pm 
\frac{\pm p \mp i (\eta \pm i\mu)}{p^2 + (\eta\pm i\mu)^2},
\end{equation}
and we obtain following expressions for an intraband contribution:
\begin{widetext}
\begin{eqnarray}
\nn
\bar\sigma^{}_{++} &=&-\frac{1}{2}\int^{}_p~{\rm Tr}\left\{\hat\sigma^{}_n {\cal P}^{}_+\hat\sigma^{}_n {\cal P}^{}_+\right\}~
\left[\frac{p + i(\eta+i\mu)}{p^2 + (\eta+i\mu)^2} - \frac{p-i(\eta-i\mu)}{p^2+(\eta-i\mu)^2}  \right]^2, 
\nn
\end{eqnarray}
and correspondingly for $\bar\sigma^{}_{--}$, as well as for an interband contribution 
\begin{eqnarray}
\nn
\bar\sigma^{}_{+-} &=&-\frac{1}{2}\int^{}_p~{\rm Tr}\left\{\hat\sigma^{}_n {\cal P}^{}_+\hat\sigma^{}_n {\cal P}^{}_-\right\}~
\left[\frac{p + i(\eta+i\mu)}{p^2 + (\eta+i\mu)^2} - \frac{p-i(\eta-i\mu)}{p^2+(\eta-i\mu)^2}\right]
\left[\frac{-p + i(\eta+i\mu)}{p^2 + (\eta+i\mu)^2} - \frac{-p-i(\eta-i\mu)}{p^2+(\eta-i\mu)^2}\right],
\end{eqnarray}
and correspondingly for $\bar\sigma^{}_{-+}$. Calculating partial traces yields:
\begin{eqnarray}
\nn
&\displaystyle{\rm Tr}\left\{\hat\sigma^{}_n {\cal P}^{}_+\hat\sigma^{}_n {\cal P}^{}_+\right\}
=\frac{p^2_n}{p^2}
\;\;\;\; {\rm and} \;\;\;\;
\displaystyle{\rm Tr}\left\{\hat\sigma^{}_n {\cal P}^{}_+\hat\sigma^{}_n {\cal P}^{}_-\right\}
=\frac{p^2_{l\neq n}}{p^2},&
\end{eqnarray}
i.e. factor $1/2$ due to angular integration. After carrying out multiplications and partial fraction decomposition in the remaining integrals we arrive at
\begin{eqnarray}
\nn
\bar\sigma^{}_{++} =\frac{1}{4}\int^{}_p~
\left[\frac{\eta}{i\mu}\left(\frac{1}{p^2+(\eta-i\mu)^2} - \frac{1}{p^2+(\eta+i\mu)^2}\right)
+\frac{2(\eta-i\mu)^2}{[p^2+(\eta-i\mu)^2]^2} + \frac{2(\eta+i\mu)^2}{[p^2+(\eta+i\mu)^2]^2}\right],
\end{eqnarray}
and
\begin{eqnarray}
\nn
\bar\sigma^{}_{+-} = \frac{\eta}{4i\mu}\int^{}_p~
\left(\frac{1}{p^2+(\eta-i\mu)^2} - \frac{1}{p^2+(\eta+i\mu)^2}\right),
\end{eqnarray}
and analogously for $\bar\sigma^{}_{--}$ and $\bar\sigma^{}_{-+}$. A straightforward integration finally yields Eqs.~(\ref{eq:InterContr}) and (\ref{eq:IntraContr}).
\end{widetext}

\section{Evaluation of diagrams in current--current formula}
\label{sec:PertCC}

\subsection{First order diagrams}

We start with the diagram $\includegraphics[height=3mm]{fig8.eps}$. For the normal channel it reads in real space representation
\begin{eqnarray}
\nn
\left.\includegraphics[height=3mm]{fig8.eps}\right|_{N}  &=& g~{\rm Tr}
\sum_{r,x}
~\sigma^{}_n G^{}_{r0}(-i\eta) \sigma^{}_n
G^{}_{0x}(i\eta) G^{}_{xx}(i\eta) G^{}_{xr}(i\eta),
\end{eqnarray}
and correspondingly for the anomalous channel
\begin{eqnarray}
\nn
\left.\includegraphics[height=3mm]{fig8.eps}\right|_{A} &=& -g~{\rm Tr}\sum_{r,x}
~\sigma^{}_n G^{}_{r0}(-i\eta) \sigma^{}_n\\
\nn
&\times& G^{}_{0x}(-i\eta) G^{}_{xx}(-i\eta) G^{}_{xr}(-i\eta).
\end{eqnarray}
Transforming into the Fourier space yields (here for both channels)
\begin{eqnarray}
\nn
&
\displaystyle
\includegraphics[height=3mm]{fig8.eps} = \pm g~{\rm Tr}\int^{}_q ~ G^{}_{q}(\pm i\eta)
&\\
\nn
&\displaystyle\times \int^{}_p ~G^{}_{p}(\pm i\eta) \sigma^{}_n G^{}_{p}(-i\eta) \sigma^{}_n G^{}_{p}(\pm i\eta), &
\end{eqnarray}
which enables one to perform integrations over momenta $q$ and $p$ separately. While the former diverges logarithmically as
$\log(\Lambda/\eta)$ with the upper cutoff $\Lambda$, the latter is zero: 
\begin{eqnarray}
\nn
\int^{}_p\frac{[\slashed p\mp i\eta][\slashed p^\dag +i\eta][\slashed p\mp i\eta]}{[p^2+\eta^2]^3}
\sim\int^\infty_0dt \frac{t-1}{[t+1]^3}=0,
\end{eqnarray}
where $t=p^2/\eta^2$, since terms containing odd powers of $p^{}_i$ and $\slashed p\slashed p^\dag$ vanish after the angular integration. For large $\Lambda$ this expression goes to zero $\sim \Lambda^{-2}$ and therefore faster than~$\log\Lambda^{-1}$. 

Next we consider diagram $\includegraphics[height=3mm]{fig6.eps}$. For both channels it reads in real space representation
\begin{eqnarray}
\nn
\includegraphics[height=3mm]{fig6.eps} &=& \pm g~{\rm Tr}\sum_{r,x}~\sigma^{}_n G^{}_{0x}(-i\eta)G^{}_{xr}(-i\eta) \\
\nn
&\times& \sigma^{}_n G^{}_{rx}(\pm i\eta) G^{}_{x0}(\pm i\eta).
\end{eqnarray} 
Changing into the Fourier representation gives
\begin{eqnarray}
\nn
\includegraphics[height=3mm]{fig6.eps} &=& \pm g~{\rm Tr}\int^{}_p ~G^{}_{p}(\pm i\eta) \sigma^{}_n G^{}_{p}(-i\eta) \\
\nn
&\times&
~\int^{}_q ~G^{}_{q}(-i\eta) \sigma^{}_n G^{}_{q}(\pm i\eta).
\end{eqnarray}
Again, both momentum integrals can be carried out separately, e.g.
\begin{eqnarray}
\nn
\int^{}_q ~G^{}_{q}(-i\eta) \sigma^{}_n G^{}_{q}(\pm i\eta) 
=\sigma^{}_n \int^{}_q\frac{[\slashed q^\dag +i\eta][\slashed q \mp i\eta]}{[q^2+\eta^2]^2} = \pm \frac{\sigma^{}_n}{4\pi}. 
\end{eqnarray}
The total correction from these diagrams is 
\begin{eqnarray}
\nn
\includegraphics[height=3mm]{fig6.eps} &=& \pm\frac{2g}{(4\pi)^2} = \pm \frac{\bar\sigma^{}_0\alpha}{4},
\end{eqnarray}
where $\alpha=g/2\pi$, and their sum is zero.

\subsection{Second order diagrams}

The evaluation of the second order diagrams without intersecting disorder lines is largely analogous to the just considered first order case. In particular, one can easily see in analogy to the diagram $\includegraphics[height=3mm]{fig8.eps}$ that diagrams $\includegraphics[height=3mm]{fig11.eps}$ and $\includegraphics[height=3mm]{fig14.eps}$ vanish as well. On the contrary, the evaluation of the diagrams with intersecting disorder lines is technically much more demanding. We start with the easier case of the diagram $\includegraphics[height=3mm]{fig13.eps}$. In real space representation, it reads (for another diagram of the class analogously)
\begin{eqnarray}
\nn
\includegraphics[height=3mm]{fig13.eps}  &=&~\pm g^2 {\rm Tr}\sum_{r,x,y}~\sigma^{}_n G^{}_{r0}(\mp i\eta) \sigma^{}_n G^{}_{0x}(i\eta) \\
\label{eq:2OrdD6}
&\times& G^{}_{xy}(i\eta)G^{}_{yx}(i\eta)G^{}_{xy}(i\eta)G^{}_{yr}(i\eta).
\end{eqnarray}
Transforming into the Fourier space and shifting momenta appropriately we arrive at
\begin{eqnarray}
\nn
&\displaystyle \includegraphics[height=3mm]{fig13.eps}  =~\pm g^2 {\rm Tr} \int^{}_k\int^{}_q~G^{}_{q+k}(i\eta)G^{}_{q}(i\eta) & \\
\nn
&\displaystyle \times\int^{}_p~G^{}_{p}(i\eta)\sigma^{}_n G^{}_{p}(\mp i\eta) \sigma^{}_n G^{}_{p}(i\eta)G^{}_{p+k}(i\eta).&
\end{eqnarray}
We start with the integral over $q$:
\begin{eqnarray}
\nn
&\displaystyle {\cal A} = \int^{}_q~G^{}_{q+k}(i\eta)G^{}_{q}(i\eta) &\\
\nn 
&\displaystyle =\int^{}_q\frac{[\slashed q+ \slashed k - i\eta][\slashed q - i\eta]}{[(k+q)^2+\eta^2][q^2+\eta^2]}.&
\end{eqnarray}
It is convenient to evaluate this integral using Feynman parametrization
\begin{equation}
\label{eq:FeynmanSimple} 
\frac{1}{AB} = \int_0^1 dx~\frac{1}{[(1-x)A+xB]^2}.
\end{equation}
Then, taking $A=q^2+\eta^2$ and symmetrizing the denominator by a momentum shift $q^{}_i\to q^{}_i-xk^{}_i$ we get 
\begin{equation}
\nn
{\cal A} = \int_0^1dx\int^{}_q~\frac{[\slashed q+(1-x)\slashed k-i\eta][\slashed q-x\slashed k-i\eta]}{[q^2+\eta^2+x(1-x)k^2]^2}.
\end{equation}
The denominator represents an even function in momenta $q$ and therefore all terms containing odd powers of $q$ in the numerator can be dropped. We get 
\begin{eqnarray}
\nn
{\cal A} &=& \int_0^1 dx \int^{}_q~
\left\{ \frac{q^2-\eta^2-(1-x)k^2}{[q^2+\eta^2+x(1-x)k^2]^2} \right.\\
\nn
&-& 
\left. \frac{i\eta(1-2x)\slashed k}{[q^2+\eta^2+x(1-x)k^2]^2} \right\}.
\end{eqnarray}
Second term in this expression does not survive under the $x$--integration. Indeed, since 
\begin{equation}
\label{eq:Xtrick} 
1-2x = \frac{d}{dx}[x(1-x)],
\end{equation}
we can substitute $x(1-x)=y$. It is possible, since $x$ enters the remaining part only via $x(1-x)$. Then, the $y$--integration runs over an empty set, since $x(1-x)=0$ for both $x=0$ and $x=1$. Integration in the first term is simple. We obtain with $\ell=\log\Lambda/\eta$ and $t=k^2/\eta^2$
\begin{eqnarray}
\nn
{\cal A} &=& \frac{1}{2\pi}\left[\ell -1\right] -\frac{1}{4\pi}\int_0^1 dx~\log[1+x(1-x)t]\\
\label{eq:D6qint}
&=& \frac{1}{2\pi}\left[\ell - \sqrt{\frac{4+t}{t}}{\rm atanh}\sqrt{\frac{t}{4+t}}\right].
\end{eqnarray}
For the evaluation of the $x$--integral it is convenient to decompose the argument of the logarithm as
\begin{equation}
\label{eq:PoleDecompos}
1+x(1-x)t = (i\sqrt{t}[x-x^{}_+])(i\sqrt{t}[x-x^{}_-]),
\end{equation}
where the poles are given by
\begin{equation}
x^{}_\pm = \frac{1}{2}\left(1 \pm \sqrt{1+ \frac{4}{t}} \right), 
\end{equation}
with the properties
\begin{equation}
1 - x^{}_{\pm} = x^{}_{\mp}, \;\; x^{}_+ + x^{}_- = 1, \;\;{\rm and}\;\; x^{}_+x^{}_- = -\frac{1}{t}.
\end{equation}
Then, the $x$--integration can be performed via the usual $\log$--integration formula $\int dx \log x = x\log x-x$.
Importantly, Eq.~(\ref{eq:D6qint}) is even under $k\to-k$ (since $t=k^2/\eta^2$). This is significant for the remaining integral over the momentum $p$, since we can also neglect terms containing odd powers of $k_i$ in the numerator after symmetrizing the denominator with respect to  $p$. In order to perform such symmetrization we employ a generalization of the Feynman--parametrization:
\begin{eqnarray}
\nn
&\displaystyle\frac{1}{A^{1+n}B} = \frac{(-1)^n}{n!}\left.\frac{\partial^n}{\partial \alpha^n}\right|_{\alpha=0}\frac{1}{[\alpha+A]B} &\\
\nn
& \displaystyle = \frac{(-1)^n}{n!}\left.\frac{\partial^n}{\partial \alpha^n}\right|_{\alpha=0} \int_0^1 dx\frac{1}{[(1-x)(\alpha+A)+xB]^2}&\\
\label{eq:FeynmanFull}
&\displaystyle =  \int_0^1dx~\frac{(n+1)(1-x)^n}{[(1-x)A+xB]^{2+n}}.&
\end{eqnarray}
The validity of the formula Eq.~(\ref{eq:FeynmanFull}) can be verified by the straightforward integration over $x$. Thus, for the $p$--integral we have 
\begin{widetext}
\begin{eqnarray}
\nn
&\displaystyle 
{\cal B} = \int^{}_p~G^{}_{p}(i\eta)\sigma^{}_n G^{}_{p}(\mp i\eta) \sigma^{}_n G^{}_{p}(i\eta)G^{}_{p+k}(i\eta)&\\ 
\nn
&
\displaystyle = 3{\rm Tr}\int_0^1dx~(1-x)^2\int^{}_p 
\frac{[\slashed p-i\eta][\slashed p^\dag\pm i\eta][\slashed p-i\eta][\slashed p+\slashed k-i \eta]}
{[(1-x)(p^2+\eta^2)+x(\eta^2+(p+k)^2)]^4}.
&
\end{eqnarray}
Shifting momentum $p^{}_i\to p^{}_i-x k^{}_i$ symmetrizes the denominator. Dropping odd powers of $p^{}_i$, $k^{}_i$, as well as $\slashed p\slashed p^\dag$ and $\slashed k\slashed k^\dag$ we get
\begin{eqnarray}
\nn
[\slashed p - x\slashed k - i\eta][\slashed p^\dag - x \slashed k^\dag \pm i\eta][\slashed p - x \slashed k-i\eta][\slashed p + (1-x)\slashed k-i \eta] 
\to \pm \eta^2[3 p^2 - \eta^2 - x(2-3x)k^2],
\end{eqnarray}
where we use the identities of the type $\slashed p\slashed k\slashed p^\dag = \slashed k^\dag p^2$. The integrals are easily performed giving
\begin{eqnarray}
\nn
&
\displaystyle
{\cal B} = \pm 3\eta^2{\rm Tr} \int_0^1dx~(1-x)^2\int^{}_p 
\frac{3 p^2 - \eta^2 - x(2-3x)k^2}{[p^2+\eta^2+x(1-x)k^2]^4}
&\\
\nn
&
\displaystyle
= \pm \frac{1}{4\pi\eta^2}\int_0^1dx~\frac{(1-x)^2(1-x(1-3x)t)}{[1+x(1-x)t]^3} = \mp \frac{1}{4\pi\eta^2}
\frac{4\sqrt{t(4 + t)} - 8(2+t){\rm atanh}\displaystyle\sqrt{\frac{t}{4+t}}}
{[t(4+t)]^{3/2}},
&
\end{eqnarray}
where again $t=k^2/\eta^2$. In order to calculate the integral over $x$ we decomposed the denominator in accord with Eq.~(\ref{eq:PoleDecompos}) and performed a partial fraction decomposition. The remaining integrals over $k$ (i.e. $t$) are innocent and can be performed numerically (for instance using Mathematica):
\begin{subequations}
\begin{eqnarray} 
&\displaystyle \mp\int_0^\infty dt~\frac{4\sqrt{[t(4+t)]}-8(2+t){\rm atanh}\displaystyle\sqrt{\frac{t}{4+t}}}
{[t(4+t)]^{3/2}} = \pm 2, &\\
\label{eq:cdef}
&
\displaystyle
\pm\int_0^\infty dt~\sqrt{\frac{4+t}{t}}{\rm atanh}\sqrt{\frac{t}{4+t}}\times\frac{4\sqrt{[t(4+t)]}-8(2+t){\rm atanh}\displaystyle\sqrt{\frac{t}{4+t}}}
{[t(4+t)]^{3/2}} \approx \mp 3.1036.
\end{eqnarray}
\end{subequations}

This yields the same conductivity correction for both channels
\begin{equation}
\label{eq:2ndD6res} 
\includegraphics[height=3mm]{fig13.eps} = \bar\sigma^{}_0 \frac{\alpha^2}{8}(2\ell-c),
\end{equation}
where $c\approx3.1036$ and $\alpha=g/2\pi$.
\end{widetext}

Next we consider diagram $\includegraphics[height=3mm]{fig15.eps}$. In real space representation it reads
\begin{eqnarray}
\nn
&
\displaystyle 
\includegraphics[height=3mm]{fig15.eps} = \pm g^2 {\rm Tr}\sum_{r,x,y}\sigma^{}_n G^{}_{ry}(\mp i\eta)G^{}_{y0}(\mp i\eta) \sigma^{}_n &\\
\nn
&
\displaystyle
\times G^{}_{0x}(i\eta)G^{}_{xy}(i\eta)G^{}_{yx}(i\eta)G^{}_{xr}(i\eta).
&
\end{eqnarray}
Fourier transforming and shifting the momenta yields
\begin{eqnarray}
\nn
&
\displaystyle 
\includegraphics[height=3mm]{fig15.eps} = \pm g^2 {\rm Tr}\int^{}_p 
\int^{}_q G^{}_{q}(\mp i\eta)\sigma^{}_n G^{}_{q}(i\eta) G^{}_{q+p}(i\eta) &\\
\nn
&
\displaystyle
\times\int^{}_k~G^{}_{k+p}(i\eta)G^{}_{k}(i\eta) \sigma^{}_n G^{}_{k}(\mp i\eta) .
&
\end{eqnarray}
We evaluate $q$--integral in details: We get rid of the current operators and use Feynman parametrization Eq.~(\ref{eq:FeynmanFull}) 
\begin{eqnarray}
\nn 
&
\displaystyle {\cal C} =
\int_q~\frac{[\slashed q^\dag \pm i\eta][\slashed q - i\eta][\slashed q + \slashed p -i\eta]}
{[q^2+\eta^2]^2[(q+p)^2+\eta^2]} = & \\
\nn
&
\displaystyle
2\int_0^1 dx~(1-x) \int^{}_q~
\frac{[\slashed q^\dag \pm i\eta][\slashed q - i\eta][\slashed q + \slashed p -i\eta]}
{[(1-x)q^2 + x(q+p)^2 + \eta^2]^3}.
&
\end{eqnarray}
Symmetrization of the denominator is achieved by shifting $q^{}_i\to q^{}_i - x p^{}_i$ such that odd powers of $q^{}_i$ in the numerator as well as terms containing $\slashed q^\dag\slashed q$ can be dropped. With some algebra in between, the numerator becomes
\begin{eqnarray}
\nn
&
\displaystyle 
[\slashed q^\dag-x\slashed p^\dag \pm i\eta][\slashed q - x \slashed p - i\eta][\slashed q +(1-x)\slashed p - i\eta]
&\\
\nn
&
\displaystyle
= (\pm i\eta - 2x\slashed p^\dag)(q^2+\eta^2 + x(1-x)p^2) 
&\\
\nn
&
\displaystyle 
- (\pm 2i\eta -3 x\slashed p^\dag)(\eta^2+x(1-x)p^2) \pm (1-2x)\eta^2\slashed p.
&
\end{eqnarray}
Terms containing $x(1-2x)$ are not shown here, since they do not survive under the $x$--integration. The integration over $q$ can now be carried out and we get
\begin{equation}
\nn
{\cal C} = -\frac{1}{4\pi}\int_0^1dx~\left[\frac{x(1-x)\slashed p^\dag}{\eta^2+x(1-x)p^2} \mp
\frac{(1-x)(1-2x)\eta^2\slashed p}{[\eta^2 + x(1-x)p^2]^2} 
\right].
\end{equation}
The $k$--integral gives the same result. We multiply both expressions, perform the trace, carry out angular integral and introduce $t=p^2/\eta^2$. The result is 
\begin{eqnarray}
\nn
\includegraphics[height=3mm]{fig15.eps}  &=& \pm\frac{2g^2}{(4\pi)^3}~\int^\infty_0\frac{dt}{t}\left(\int_0^1 dx~\frac{x(1-x)t}{1+x(1-x)t} \right)^2 \\
\nn
&\pm &
\frac{2g^2}{(4\pi)^3}~\int^\infty_0\frac{dt}{t}\left(\int_0^1dx~\frac{(1-x)(1-2x)t}{[1+x(1-x)t]^2} \right)^2.
\end{eqnarray}
Second $x$--integral be reduced to the first by an integration by parts:
\begin{eqnarray}
\nn
&\displaystyle \int_0^1dx~\frac{(1-x)(1-2x)t}{[1+x(1-x)t]^2} &\\
\nn
&\displaystyle = - \int_0^1dx~(1-x)\frac{\partial}{\partial x}\frac{1}{1+x(1-x)t} &\\
\nn
&\displaystyle
= 1 - \int_0^1 dx~\frac{1}{1+x(1-x)t} 
&\\
\nn
&
\displaystyle = \int_0^1 dx~\frac{x(1-x)t}{1+x(1-x)t}.
&
\end{eqnarray}
The $x$--integral can be easily calculated, which yields for the diagram 
\begin{equation}
\label{eq:2ndOrdD8resA} 
\includegraphics[height=3mm]{fig15.eps}  = \pm \frac{4g^2}{(4\pi)^3}\int_0^\infty\frac{dt}{t}\left(1 - \frac{4}{t}\sqrt{\frac{t}{4+t}}{\rm atanh}\sqrt{\frac{t}{4+t}} \right)^2.
\end{equation}
The expression in the brackets behaves well for all $t$. For $t\to0$ it goes to $t/6$ and therefore the integral is innocent at the lower integration boundary. For $t\to\infty$ it approaches unity and the integral diverges logarithmically,
\begin{equation}
\int^{e^{2\ell}} \frac{dt}{t} \sim 2\ell,
\end{equation}
since $t\propto k^2$. Numerical evaluation gives 
\begin{equation}
\nn
\int_0^{e^{2\ell}}~\frac{dt}{t}~\left(1 - \frac{4}{t}\sqrt{\frac{t}{4+t}}{\rm atanh}\sqrt{\frac{t}{4+t}}\right)^2 = 2\ell - c,
\end{equation}
where $c\approx3.1036$. The conductivity contribution is
\begin{equation}
\label{eq:2ndOrdD8resB} 
\includegraphics[height=3mm]{fig15.eps}  = \pm \bar\sigma^{}_0\frac{\alpha^2}{4}(2\ell-c),
\end{equation}
and gives zero by summing over both channels.

The last diagram $\includegraphics[height=3mm]{fig10.eps}$ reads in Fourier representation
\begin{eqnarray}
\nn 
&
\displaystyle
\includegraphics[height=3mm]{fig10.eps}  = \pm g^2 {\rm Tr}\int^{}_q
\int^{}_p~G^{}_p(\mp i\eta)\sigma^{}_n G^{}_p(i\eta)G^{}_{p+q}(i\eta)&\\
\nn
&
\displaystyle 
\times\int^{}_k~G^{}_{k}(i\eta)\sigma^{}_n G^{}_{k}(\mp i\eta) G^{}_{k-q}(\mp i\eta),
&
\end{eqnarray}
which resembles strongly the just considered diagram $\includegraphics[height=3mm]{fig15.eps}$. The evaluation goes analogously with some minor changes. For instance, while evaluating integral over $k$, the momenta should be shifted as $k^{}_i\to k^{}_i +x q^{}_i$. The result has the same numerical value with the same sign in  both channels:
\begin{equation}
\label{eq:2ndOrdD3} 
\includegraphics[height=3mm]{fig10.eps}  = \bar\sigma^{}_0\frac{\alpha^2}{4}(2\ell-c).
\end{equation}

\begin{widetext}
\section{Density--density formula}
\label{sec:BandsDD}

\subsection{Band contributions}
Making use of the scaling relation Eq.~(\ref{eq:scaling}) we write the Kubo formula for dc conductivity in the density--density formula with an artificial replacement of the frequency by some small but finite quantity $\eta$ usually associated with the inverse scattering time:
\begin{eqnarray}
\nn
\bar\sigma &=& 2\eta^2{\rm Tr}\sum_r r^2_n G^{}_{0r}(-\mu+i\eta) G^{}_{r0}(-\mu-i\eta) 
=-2\eta^2\left.\frac{\partial^2}{\partial q^2_n}\right|_{q=0} {\rm Tr}\int_p~G^{}_p(-\mu+i\eta) G^{}_{p+q} (-\mu-i\eta)\\
\label{eq:KuboDD1} 
&=& 2\eta^2 {\rm Tr}\int_p ~G^{}_p(- \mu + i\eta )\sigma^{}_n G^{}_p(- \mu+i\eta ) 
G^{}_p(- \mu-i\eta )\sigma^{}_n G^{}_p(- \mu-i\eta ),
\end{eqnarray}
where in the last line we performed a partial integration and used then the usual differentiation formula for matrices $\partial^{}_x A^{-1} = - A^{-1}~\partial^{}_x A~A^{-1}$. Next we go into the diagonal representation and project out. The $\bar\sigma^{}_{++}$ and $\bar\sigma^{}_{+-}$ contributions then read
\begin{eqnarray} 
\bar\sigma^{}_{++} &=& 2\eta^2{\rm Tr}\int_p~\hat G^{}_p(-\mu+i\eta)\hat\sigma^{}_n{\cal P}^{}_+\hat G^{}_p(-\mu+i\eta)
\hat G^{}_p(-\mu-i\eta)\hat\sigma^{}_n{\cal P}^{}_+\hat G^{}_p(-\mu-i\eta)\\
\bar\sigma^{}_{+-} &=& 2\eta^2{\rm Tr}\int_p~\hat G^{}_p(-\mu+i\eta)\hat\sigma^{}_n{\cal P}^{}_+\hat G^{}_p(-\mu+i\eta)
\hat G^{}_p(-\mu-i\eta)\hat\sigma^{}_n{\cal P}^{}_-\hat G^{}_p(-\mu-i\eta)
\end{eqnarray}
We perform the trace and angular integration:
\begin{eqnarray}
\bar\sigma^{}_{++} &=& 
\frac{\eta^2}{2\pi}\int_0^\infty dp p~
\frac{[p+i(\eta-i\mu)]^2[p-i(\eta+i\mu)]^2}
{[p^2+(\eta-i\mu)^2]^2[p^2+(\eta+i\mu)^2]^2} = \frac{\bar\sigma^{}_0}{4}(1+z~{\rm atan}(z)),\\
\bar\sigma^{}_{+-} &=& 
\frac{\eta^2}{2\pi}\int_0^\infty dp p~
\frac{1}
{[p^2+(\eta-i\mu)^2][p^2+(\eta+i\mu)^2]}  = \frac{\bar\sigma^{}_0}{4z}~{\rm atan}(z).
\end{eqnarray}
Counting all contributions together gives Eq.~(\ref{eq:Cond}).

\subsection{First order perturbation theory at half filling}

Consider disorder averaged conductivity for the case of scalar disorder
\begin{equation}
\label{eq:Kubo6}
\bar\sigma = -2\eta^2\left.\frac{\partial^2}{\partial q^2_n}\right|_{q=0}{\rm Tr}\int_p\langle G^{}_p(i\eta)G^{}_{p+q}(-i\eta)\rangle,
\end{equation}
where $\langle\cdot\cdot\rangle$ denotes averaging with respect to disorder. Performing first perturbative expansion and then derivatives with respect to $q^{}_n$ we arrive at
\begin{subequations}
\begin{eqnarray}
\label{eq:1stPTa}
\bar\sigma &=& -4g\eta^2{\rm Tr}\int_k~G^{}_k(i\eta)~\int_p~G^{}_p(i\eta)G^{}_p(-i\eta)\sigma^{}_n G^{}_p(-i\eta)\sigma^{}_n G^{}_p(-i\eta) G^{}_p(i\eta) \\
\label{eq:1stPTb}
&& -4g\eta^2{\rm Tr}\int_k~G^{}_k(-i\eta)~\int_p~G^{}_p(-i\eta)G^{}_p(i\eta)\sigma^{}_n G^{}_p(i\eta)\sigma^{}_n G^{}_p(i\eta) G^{}_p(-i\eta)\\
\label{eq:1stPTc}
&& -4g\eta^2{\rm Tr}\int_k~G^{}_k(i\eta)G^{}_k(-i\eta)~\int_p~G^{}_p(-i\eta)\sigma^{}_n G^{}_p(-i\eta)\sigma^{}_n G^{}_p(-i\eta) G^{}_p(i\eta) \\
\label{eq:1stPTd}
&& -4g\eta^2{\rm Tr}\int_k~G^{}_k(-i\eta)G^{}_k(i\eta)~\int_p~G^{}_p(i\eta)G^{}_p(-i\eta)\sigma^{}_n G^{}_p(-i\eta)\sigma^{}_n G^{}_p(-i\eta) \\
\label{eq:1stPTe}
&&  -4g\eta^2{\rm Tr}\int_k~G^{}_k(-i\eta)\sigma^{}_n G^{}_k(-i\eta)G^{}_k(i\eta)~\int_p~G^{}_p(i\eta)G^{}_p(-i\eta)\sigma^{}_n G^{}_p(-i\eta).
\end{eqnarray}
\end{subequations}
Eqs.~(\ref{eq:1stPTa}) and (\ref{eq:1stPTb}) correspond to both $\includegraphics[height=3mm]{fig8.eps}$--diagrams, while Eqs.~(\ref{eq:1stPTc}), (\ref{eq:1stPTd}) and (\ref{eq:1stPTd}) together represent the contribution from the diagram $\includegraphics[height=3mm]{fig6.eps}$. Bearing in mind that
$$
G^{}_p(i\eta)G^{}_p(-i\eta) = \frac{1}{p^2+\eta^2}
\,\,\,\,{\rm and } \,\,\,\,
\int_p~G^{}_p(\pm i\eta) = \mp\frac{\ell}{2\pi},
$$
we immediately obtain for each contribution
\begin{eqnarray}
\label{eq:ct1}
{\rm Eq.~(\ref{eq:1stPTa})} &=& {\rm Eq.~(\ref{eq:1stPTb})} = - \bar\sigma^{}_0\alpha\ell,\\
\label{eq:ct2}
{\rm Eq.~(\ref{eq:1stPTc})} &=& {\rm Eq.~(\ref{eq:1stPTd})} =   \bar\sigma^{}_0\alpha\ell,\\
\label{eq:ct3}
{\rm Eq.~(\ref{eq:1stPTe})} &=& \bar\sigma^{}_0\alpha.
\end{eqnarray}
Evidently, the sum of all contributions is free of logarithms but finite in contrast to the same order perturbative correction from the current--current formula, where it is zero. 

\subsection{First order perturbation theory at nonzero chemical potential}

We start with diagrams $\includegraphics[height=3mm]{fig8.eps}$ and perform first the derivatives:
\begin{eqnarray} 
\nn
&\displaystyle
\sum_{\rm all} \includegraphics[height=3mm]{fig8.eps}  = -2g\eta^2 \left.\frac{\partial^2}{\partial q^2_n}\right|_{q=0}~
{\rm Tr}\sum_{s=\pm} \int^{}_k G^{}_k(-\mu+is\eta)~\int^{}_p G^{}_{p}(-\mu+is\eta)G^{}_{p+q}(-\mu-is\eta)G^{}_p(-\mu+is\eta)
& \\
\nn
&\displaystyle
= -4g\eta^2 {\rm Tr}\sum_{s=\pm}\int^{}_k G^{}_k(-\mu+is\eta)\int^{}_p G^{}_p(-\mu+is\eta) G^{}_{p}(-\mu-is\eta)\sigma^{}_n 
G^{}_{p}(-\mu-is\eta)\sigma^{}_n G^{}_{p}(-\mu-is\eta)G^{}_p(-\mu+is\eta).
&
\end{eqnarray}
Using the notation $A=\eta - is\mu$ and $B=\eta + is\mu$ we obtain for $k$--integral
\begin{equation}
\int_k G^{}_k(-\mu+is\eta) = -is\frac{A}{4\pi}\log\frac{\Lambda^2}{A^2},
\end{equation}
and for $p$--integral 
\begin{eqnarray}
\nn
&\displaystyle
{\rm Tr}\int^{}_p G^{}_p(-\mu+is\eta) G^{}_{p}(-\mu-is\eta)\sigma^{}_n 
G^{}_{p}(-\mu-is\eta)\sigma^{}_n G^{}_{p}(-\mu-is\eta)G^{}_p(-\mu+is\eta) &\\
\nn
&\displaystyle =
is\frac{B}{2\pi}\int_0^{\infty}dt~\frac{t^2 + t(4AB-A^2-B^2)+A^2B^2}{[t+A^2]^2[t+B^2]^3}.
&
\end{eqnarray}
Performing the partial fraction decomposition with the coefficients 
\begin{eqnarray}
\nn
\alpha^{}_1 = \frac{2AB-3A^2-B^2}{(A-B)^2(A+B)^4}, & & \beta^{}_2 = -\frac{A-B}{(A+B)^3},\\
\nn
\alpha^{}_2 = - \frac{2A^2}{(A-B)(A+B)^3}, & & \beta^{}_3 = \frac{2B^2}{(A+B)^2},
\end{eqnarray}
we obtain for the conductivity correction from the $\includegraphics[height=3mm]{fig8.eps}$--diagram class
\begin{eqnarray} 
\nn
\displaystyle
\sum_{\rm all} \includegraphics[height=3mm]{fig8.eps}  = -\bar\sigma^{}_0\alpha\eta^2 \sum_{s=\pm} AB\log\frac{\Lambda^2}{A^2}
\left( 
\alpha^{}_1\log\frac{B^2}{A^2} + \frac{\alpha^{}_2}{A^2} + \frac{\beta^{}_2}{B^2} + \frac{1}{2}\frac{\beta^{}_3}{B^4}
\right).
\end{eqnarray}
Since $A+B=2\eta$ the factor $\eta^2$ in front drops out. Using 
\begin{equation}
\log\frac{\Lambda^2}{A^2} = \log\frac{\Lambda^2}{\mu^2+\eta^2} + 2is~{\rm atan}\left(\frac{\mu}{\eta}\right) \;\;
{\rm and} \;\;
\log\frac{B^2}{A^2} = 4is~{\rm atan} \left(\frac{\mu}{\eta}\right),
\end{equation}
eventually leads us to the following expression 
\begin{eqnarray}
\label{eq:contrII}
\sum_{\rm all} \includegraphics[height=3mm]{fig8.eps} &=& -\frac{\bar\sigma^{}_0\alpha}{2} \left[ 
\ell\left(1+ \frac{1+z^2}{z}{\rm atan}(z)\right) + \frac{1+3z^2}{2z}{\rm atan}(z) - \frac{1 - 2z^2 - 3z^4}{2z^2} {\rm atan}^2(z)
\right]
\end{eqnarray}
with $\displaystyle 2\ell = \log\frac{\Lambda^2}{\eta^2+\mu^2}$. As $z\to0$ we reproduce Eq.~(\ref{eq:ct1}).

Next we evaluate the diagram $\includegraphics[height=3mm]{fig6.eps}$. Upon taking the derivatives with respect to $q_n$ we have
\begin{subequations}
\begin{eqnarray}
\nn
&\displaystyle
{\includegraphics[height=3mm]{fig6.eps}} = -2g\eta^2 \left.\frac{\partial^2}{\partial q^2_n}\right|_{q=0}~{\rm Tr}\int^{}_p
G^{}_p(-\mu+i\eta)G^{}_{p+q}(-\mu-i\eta)\int^{}_kG^{}_{k+q}(-\mu-i\eta)G^{}_k(-\mu+i\eta) &\\
\label{eq:CTm1}
&\displaystyle
= -4g\eta^2{\rm Tr}\int^{}_p G^{}_p(-\mu+i\eta)G^{}_p(-\mu-i\eta)\sigma^{}_nG^{}_p(-\mu-i\eta)\sigma^{}_nG^{}_p(-\mu-i\eta)
\int^{}_k G^{}_k(-\mu-i\eta)G^{}_k(-\mu+i\eta)
&\\
\label{eq:CTm2}
&\displaystyle
-4g\eta^2{\rm Tr}\int^{}_p G^{}_p(-\mu+i\eta)G^{}_p(-\mu-i\eta)
\int^{}_kG^{}_k(-\mu-i\eta)\sigma^{}_nG^{}_{k}(-\mu-i\eta)\sigma^{}_nG^{}_k(-\mu-i\eta)G^{}_k(-\mu+i\eta) 
&\\
\label{eq:CTm3}
&\displaystyle
 -4g\eta^2 {\rm Tr}\int^{}_p G^{}_p(-\mu+i\eta)G^{}_p(-\mu-i\eta)\sigma^{}_nG^{}_p(-\mu-i\eta)
\int^{}_k G^{}_k(-\mu-i\eta)\sigma^{}_n G^{}_k(-\mu-i\eta)G^{}_k(-\mu+i\eta).
&
\end{eqnarray}
\end{subequations}
Obviously, Eqs.~(\ref{eq:CTm1}--\ref{eq:CTm3}) reduce to Eqs.(\ref{eq:1stPTc}--\ref{eq:1stPTe}) as $\mu\to0$. Eqs.~(\ref{eq:CTm1}) and (\ref{eq:CTm2}) give identical contributions. 
Both integrals can be carried out separately. After performing the trace we get
\begin{equation}
-16g\eta^2\int\frac{d^2k}{(2\pi)^2}~\frac{k^2+\mu^2+\eta^2}{[k^2+(\eta+i\mu)^2][k^2+(\eta-i\mu)^2]}~
\int\frac{d^2p}{(2\pi)^2}~\frac{i(\eta+i\mu)[i(\eta+i\mu)[p^2+\mu^2+\eta^2]-2\mu p^2]}
{[p^2+(\eta-i\mu)^2][p^2+(\eta+i\mu)^2]^3}
\end{equation}
Introducing $A=\eta-i\mu$ and $B=\eta+i\mu$ we get by the partial fraction decomposition
\begin{eqnarray}
\bar\sigma^{}_0\alpha\eta^2B\int_0^\infty dx\left(\frac{\alpha}{x+A^2} + \frac{\beta}{x+B^2}\right)
\int_0^\infty dy~\left[a_1\left(\frac{1}{y+A^2} - \frac{1}{y+B^2}\right) + \frac{b_2}{(y+B^2)^2} + \frac{b_3}{(y+B^2)^3}\right]
\end{eqnarray}
with
\begin{eqnarray}
\nn
\alpha = \frac{A}{A+B},\;\; \beta = \frac{B}{A+B},\;\;
a^{}_1 = -\frac{A}{(A-B)(A+B)^3}, \;\; b^{}_2 = -\frac{A}{(A+B)^2},\;\; { \rm and} \;\; b^{}_3 = \frac{2B^2}{A+B}. 
\end{eqnarray}
The integration over $x$ yields 
$$
2(\ell +z~{\rm atan}(z)),
$$
while $y$--integration gives after multiplication with $B$
$$
\frac{1}{4\eta^2}\left(1+ \frac{1+z^2}{z}{\rm atan}(z)\right)
$$
Piecing all terms together gives for the contribution
\begin{equation}
{\rm Eq.(\ref{eq:CTm1})+Eq.(\ref{eq:CTm2})} = \frac{\bar\sigma^{}_0\alpha}{2} (\ell +z~{\rm atan}(z)) \left(1+ \frac{1+z^2}{z}{\rm atan}(z)\right).
\end{equation}
Evidently, terms which contain $\ell$ are precisely the same as whose appearing in $\includegraphics[height=3mm]{fig8.eps}$ in Eq.~(\ref{eq:contrII}) but with the opposite sign. 
Therefore, the $\ell$--dependent terms will have gone after summing them. Eq.~(\ref{eq:CTm3}) does not develop any logarithms. This contribution can be written after performing the trace as
\begin{equation}
\frac{1}{2}\bar\sigma^{}_0\alpha\eta^2\left(A\int_0^\infty dx~\frac{AB+x}{[x+A^2][x+B^2]^2} \right)^2 =  
\frac{1}{8}\bar\sigma^{}_0\alpha\left(1 + \frac{1+z^2}{z}{\rm atan}(z) \right)^2.
\end{equation}
Therefore, the total first order conductivity correction in density--density formula reads
\begin{equation}
\label{eq:1stCorrDD}
{\includegraphics[height=3mm]{fig6.eps}}+\sum_{\rm all}{\includegraphics[height=3mm]{fig8.eps}} =
\frac{\bar\sigma^{}_0\alpha}{8}\left(1+\frac{3+2z^2-z^4}{z^2}{\rm atan}^2(z)\right),
\end{equation}
with the both limits $\bar\sigma^{}_0\alpha/2$ as $z\to0$ and $-\bar\sigma\alpha\pi^2 z^2/32$ as $z\to\infty$.

\end{widetext}


\begin{thebibliography}{99}
%
\bibitem{Gorkov1979} L. G. Gor'kov, A. I. Larkin, and D. E. Khmel'nitskii, Pis'ma Zh. Eksp. Teor. Fiz. {\bf 30}, 248 (1979) [JETP Lett. {\bf 30}, 228 (1979)].
%
\bibitem{Altshuler1980}
B. L. Altshuler, D. E. Khmel'nitskii, A. I. Larkin, and P. A. Lee, Phys. Rev. B {\bf 22}, 5142 (1980);
B. L. Altshuler and B. D. Simons, in {\it Mesoscopic quantum physics},
eds. E. Akkermans et al., North-Holland Publishing, Amsterdam (1995).
%
\bibitem{Hikami1980} S. Hikami, A. Larkin, and Y. Nagaoka, Prog. Theor. Phys. {\bf 63}, 707 (1980).
%
\bibitem{Altshuler1981} B. L. Altshuler, A. G. Aronov, A. I. Larkin, and D. E. Khmel'nitskii, Th. Eksp. Teor. Fiz. {\bf 81}, 768 (1981) [Sov. Phys. JETP {\bf 54}, 411 (1981)].
%
\bibitem{Ando2002}
T. Ando, Y. Zheng and H. Suzuura, J. Phys. Soc. Japan {\bf 71}, 1318 (2002).
%
\bibitem{McCann2006} E. McCann and V. I. Falko, Phys. Rev. Lett. {\bf 96} 086805 (2006); 
E. McCann, K. Kechedzhi, V. I. Falko, H. Suzuura, T. Ando, and B. L. Altshuler, {\it ibid} {\bf 97}, 146805 (2006).
%
\bibitem{Suzuura2002} H. Suzuura and T. Ando, Phys. Rev. Lett. {\bf 89}, 266603 (2002).
%
\bibitem{Khveshchenko2006} D. V. Khveshchenko, Phys. Rev. Lett. {\bf 97}, 036802 (2006).
%
\bibitem{Wegner1979} F. Wegner, Z. Physik B {\bf 35}, 207 (1979).
%
\bibitem{Wegner1980} L. Sch\"a{}fer and F. Wegner, Z. Physik B {\bf 38}, 113 (1980).
%
\bibitem{Wegner1979a} F. J. Wegner, Phys. Rev. B {\bf 19},  783 (1979).
%
\bibitem{McKane1981} A. J. McKane and M. Stone, Ann. Phys. {\bf 131}, 36 (1981).
%
\bibitem{Hikami1981} S. Hikami, Phys. Rev. B {\bf 24}, 2671 (1981).
%
\bibitem{Efetov1983} K. B. Efetov, Adv. Phys. {\bf 32}, 53 (1983); 
K. B. Efetov, {\it Supersymmetry in Disorder and Chaos},  Cambridge University Press, New York, (1997).
%
\bibitem{Fradkin1986} E. Fradkin, Phys. Rev. {\bf 33}, 3257 (1986);  {\it ibid} 3263 (1986).
%
\bibitem{Wegner1989} F. Wegner, Nucl. Phys. B {\bf 316}, 663 (1989).
%
\bibitem{ziegler07} K. Ziegler, Phys. Rev. B {\bf 75}, 233407 (2007).
%
\bibitem{Ziegler2009} K. Ziegler, Phys. Rev. Lett. {\bf 102}, 126802 (2009); 
Phys. Rev. B {\bf 79}, 195424 (2009).
%
\bibitem{Kubo} R. Kubo, M. Troda, and N. Hashitsume, {\it Statistical Physics}, Vol. II, Springer, Berlin (1992).
%
\bibitem{Ziegler2008} K. Ziegler, Phys. Rev. B {\bf 78}, 125401 (2008).
%
\bibitem{ziegler12} K. Ziegler,  Eur. Phys. J. B {\bf 86}, 391 (2013).
%
\bibitem{Novoselov2005} K. S. Novoselov, A. K. Geim, S. V. Morozov, D. Jiang, 
M. I. Katsnelson, I. V. Grigorieva, S. V. Dubonos, and A. A. Firsov, Nature (London) {\bf 438}, 197 (2005).
%
\bibitem{Sinner2011} A. Sinner and K. Ziegler, Phys. Rev. B {\bf 84}, 233401 (2011).
%
\bibitem{Aleiner2006} I. Aleiner and K. Efetov, Phys. Rev. Lett. {\bf 97} 236801 (2006).
%
\bibitem{Ziegler1997} K. Ziegler, Phys. Rev. B {\bf 55}, 10661 (1997); K. Ziegler and G. Jug, Z. Phys. B {\bf 104}, 5 (1997); K. Ziegler, Phys. Rev. Lett. {\bf 80}, 3113 (1998).
%
\bibitem{Bocquet2000} M. Bocquet, D. Serban, and M. R. Zirnbauer, Nucl. Phys. B {\bf 578}, 628 (2000).
%
\bibitem{Bardarson2010} J. H. Bardarson, M. V. Medvedyeva, J. Tworzyd\l{}o, A. R. Akhmerov, and C. W. J. Beenakker, 
Phys. Rev. B {\bf 81}, 121414(R) (2010).
%
\bibitem{Medvedeva2010} M. V. Medvedyeva, J. Tworzyd\l{}o, and C. W. J. Beenakker, Phys. Rev. B {\bf 81}, 214203 (2010). 
%
\bibitem{Ludwig1994} A. W. W. Ludwig, M. P. A. Fisher, R. Shankar, and G. Grinstein,
Phys. Rev. B {\bf 50}, 7526 (1994).
%
\bibitem{Bhaseen2000} M. J. Bhaseen, I. I. Kogan, O. A. Soloviev, N. Taniguchi, A.Tsvelik, Nucl. Phys. B {\bf 580}, 688 (2000);
M. J. Bhaseen, J.--S. Caux, I. I. Kogan, A. M. Tsvelik, {\it ibid} B {\bf 618}, 465 (2001).
%
\bibitem{ziegler13}
K. Ziegler and A. Sinner, arXiv:1311.3441.
%
\bibitem{Ziegler2012} K. Ziegler, J. Phys. A: Math. Theor. {\bf 45}, 335001 (2012). 
%
\end{thebibliography}
\end{document}